\documentclass[journal, 10pt]{IEEEtran}

\usepackage{amsfonts, amsmath, amsthm,amscd,mathrsfs} 
\usepackage{fixmath}

\usepackage[hidelinks]{hyperref}

\def\ds{\displaystyle}

\newcommand{\bbE}{\mathbb{E}}
\newcommand{\bbR}{\mathbb{R}}
\newcommand{\bbZ}{\mathbb{Z}}

\newcommand{\bfJ}{\mathbf{J}}

\newcommand{\sff}{\mathsf{f}}

\newcommand{\rmd}{\mathrm{d}}
\newcommand{\rme}{\mathrm{e}}

\newcommand{\cI}{\mathcal{I}}
\newcommand{\cL}{\mathcal{L}}
\newcommand{\cM}{\mathcal{M}}
\newcommand{\cN}{\mathcal{N}}
\newcommand{\cR}{\mathcal{R}}
\newcommand{\cY}{\mathcal{Y}}

\newcommand{\mfrakS}{\mathfrak{S}}
\newcommand{\mfrakT}{\mathfrak{T}}
\newcommand{\mfrakV}{\mathfrak{V}}
\newcommand{\mfrakW}{\mathfrak{W}}

\usepackage{thmtools}

\theoremstyle{plain}
\newtheorem{thm}{Theorem}%[section]
\theoremstyle{plain}
\newtheorem{lem}{Lemma}%[section]
\theoremstyle{plain}
\newtheorem{prop}{Proposition}%[section]

\theoremstyle{plain}
\newtheorem*{exa*}{Example}%[section]

\makeatletter
\def\thmhead@plain#1#2#3{%
  \thmname{#1}\thmnumber{\@ifnotempty{#1}{ }\@upn{#2}}%
  \thmnote{ {\the\thm@notefont#3}}}
\let\thmhead\thmhead@plain
\makeatother

\usepackage{scrextend}
\usepackage{cite}

\newcommand{\boldtheta}{\mathbold{\theta}}
\newcommand{\boldtau}{\mathbold{\tau}}
\newcommand{\boldalpha}{\mathbold{\alpha}}
\newcommand{\boldxi}{\mathbold{\xi}}
\newcommand{\bolda}{\mathbold{a}}
\newcommand{\boldt}{\mathbold{t}}

\newcommand{\opD}{\operatorname{D}}
\newcommand{\opGamma}{\operatorname{\Gamma}}
\newcommand{\diag}{\operatorname{diag}}

\newcommand{\simplex}{\operatorname{\Delta}}

\usepackage{mathtools}
\makeatletter
\newcommand{\vast}{\bBigg@{4}}
\newcommand{\Vast}{\bBigg@{5}}
\newcommand{\Gigantic}{\bBigg@{8}}

\makeatother

\makeatletter
\newcommand\footnoteref[1]{\protected@xdef\@thefnmark{\ref{#1}}\@footnotemark}
\makeatother

\DeclareMathOperator*{\esssup}{ess\,sup}

\allowdisplaybreaks

\begin{document}
\title{Minimax R\'enyi Redundancy}
\author{Semih Yagli,~\IEEEmembership{Student Member, IEEE}, Y\"{u}cel Altu\u{g}, and Sergio Verd\'u,~\IEEEmembership{Fellow, IEEE}
\thanks{Semih Yagli and Sergio Verd\'u are with the Electrical Engineering Department, Princeton University, Princeton, NJ 08544. Y\"ucel Altu\u{g} is with Natera Inc., San Carlos, CA 94070. E-mail: {\em syagli@princeton.edu}, {\em yucelaltug@gmail.com}, {\em verdu@princeton.edu}. Part of this paper was presented at 2017 IEEE International Symposium on Information Theory \cite{Yagli2017ISIT}.
}}
\maketitle

\begin{abstract}

 The redundancy for universal lossless compression of discrete memoryless sources in Campbell's setting is characterized as a minimax R\'enyi divergence, which is shown to be equal to the maximal $\alpha$-mutual information via a generalized redundancy-capacity theorem. Special attention is placed on the analysis of the asymptotics of minimax R\'enyi divergence, which is determined up to a term vanishing in blocklength.


\end{abstract}

\section{Introduction} \label{sec:introduction}

\IEEEPARstart{I}{n} variable length source coding, expected code length is the usual cost function that one aims to minimize. For discrete memoryless sources, asymptotically, the minimal achievable per-letter expected code length is equal to the entropy. However, if $P_{Y^n|V=\boldtheta }$ is a discrete memoryless source distribution with an unknown parameter $\boldtheta$ and the encoding system assumes a distribution $Q_{Y^n}$, then one needs to pay an extra penalty for the mismatch given by\footnote{For prefix codes, \eqref{eqn:redundancy} is well known \cite[Theorem 5.4.3]{CoverThomas}. On the other hand, the loss in rate incurred due to the prefix condition is known to be asymptotically negligible \cite{KontoyiannisVerdu2014}.}
\begin{align}
	\frac{1}{n} D(P_{Y^n|V=\boldtheta}\| Q_{Y^n} )+ o(1)\text{,} \label{eqn:redundancy}
\end{align}
where $D(P\|Q)$ stands for the relative entropy between the probability measures $P$ and $Q$. In light of \eqref{eqn:redundancy}, the conventional worst-case measure of redundancy in universal lossless compression is
\begin{align}
	R_n=\inf_{Q_{Y^n} } \sup_{\boldtheta }D(P_{Y^n|V=\boldtheta }\| Q_{Y^n})\text{,} \label{eqn:minimax redundancy definition}
\end{align}
where the infimization is over all distributions on $\cY^n$, and the supremum is over all possible values of the unknown parameter. In this zero-sum game, $Q_{Y^n}$ is chosen by the code designer, and $\boldtheta$ is chosen by nature.

A relation between $R_n$ and the maximal mutual information is given by the \emph{Redundancy-Capacity Theorem} (e.g., \cite{Gallagher1976}, and \cite{Ryabko1979}) that states that
\begin{align}
	R_n=\sup_{P_V} I(P_V, P_{Y^n|V})\text{,} \label{eqn:redundancy-capacity theorem}
\end{align}
where\footnote{$I(P_X, P_{Y|X}) = D( P_{Y|X}P_X \| P_Y P_X )$ is the mutual information between $X$ and $Y$ with $(X,Y)\sim P_X P_{Y|X}$.} the supremization is over all probability distributions on the parameter space. Through \eqref{eqn:redundancy}, \eqref{eqn:minimax redundancy definition} and \eqref{eqn:redundancy-capacity theorem}, we see a pleasing relationship between entropy, relative entropy and mutual information in the context of lossless data compression.

Let $Y^n\sim P_{Y^n|V=\boldtheta}$, and note that
\begin{align}
	D(P_{Y^n|V=\boldtheta}\| Q_{Y^n} )=\bbE \left[\imath_{P_{Y^n|V=\boldtheta}\| Q_{Y^n} }(Y^n) \right]\text{,} 
\end{align}
where the \emph{relative information} between the discrete probability measures $P$ and $Q$ is defined as\footnote{Unless otherwise stated, logarithms and exponentials are of arbitrary basis.}
\begin{align}
\imath_{P\| Q }(a)=\log \frac{P(a)}{Q(a)}\text{.}
\end{align}
A much more stringent performance guarantee than the average of relative information is its pointwise maximum. In particular, if one replaces $\bbE[\imath_{P_{Y^n|V=\boldtheta}\| Q_{Y^n} }(Y^n) ]$ with $\max_{y^n} \imath_{P_{Y^n|V=\boldtheta}\| Q_{Y^n} }(y^n) $ in \eqref{eqn:minimax redundancy definition}, the resulting quantity, i.e.,
\begin{align}
	r_n=\inf_{Q_{Y^n} } \sup_{\boldtheta } \max_{y^n\in \cY^n} \imath_{P_{Y^n|V=\boldtheta}\| Q_{Y^n} }(y^n)\text{,} \label{eqn:minimax regret definition}
\end{align}
is called the \emph{minimax regret}, which has found applications in various settings\footnote{For example, in lossless compression with prefix codes, $\imath_{P_{Y^n|V=\boldtheta}\| Q_{Y^n} }(y^n)$ is often viewed as a proxy for the mismatch penalty incurred by assuming that $y^n$ is drawn from $Q_{Y^n}$ rather than the true distribution $P_{Y^n|V=\boldtheta}$. Such an approximation can be justified asymptotically.}, e.g., \cite{XieBarron2000, Shtarkov87, ForsterWarmuth2002, Dramota04, LiangBarron2004}. An analogy to the Redundancy-Capacity Theorem is given by \cite{Shtarkov87}
\begin{align}
r_n&= \log \sum_{y^n\in \cY^n} \sup_{\boldtheta} P_{Y|V=\boldtheta}(y^n)  \label{eqn:shtarkov sum} \\
&=\sup_{P_V}I_\infty(P_V, P_{Y^n|V})\text{,} \label{eqn:redundancy-capacity for Regret}
\end{align}
where $I_\infty(P_X, P_{Y|X})$ denotes the $\alpha$-mutual information of infinite order, whose definition is given in \eqref{eqn:alpha-mutual-information definition}.

The average and pointwise formulations are two extremes of performance guarantees, which are not quite suitable for certain applications. For this reason, one seeks a compromise between those two. For example, in the economics literature, average and pointwise guarantees are referred as \emph{risk-neutral} and \emph{risk-avoiding}, respectively. Since the former is known to be too lenient and the latter is known to be too stringent for typical applications, the notion of \emph{risk-aversion} has been introduced to provide a more useful compromise between these two extremes \cite{Pratt64}, \cite{Arrow65}, which is known to be relevant for diverse applications \cite{Ross81}. In this paper, we introduce the notion of risk-aversion within the universal source coding context and quantify its effect on the fundamental limit.

In the non-universal setting, i.e., when the source distribution is known, a classical result of Campbell \cite{Campbell1965} introduces such a risk-averse cost function in a discrete memoryless setting. Specifically, \cite{Campbell1965} proposes to generalize the conventional notion of minimizing the expected code length with the cost function
\begin{align}
	L_\lambda(Y^n)=\frac{1}{\lambda}\log\bbE \left[ \exp (\lambda \ell(\sff(Y^n) ) ) \right]\text{,}
\end{align}
where $\lambda\in (0, \infty) $, $\sff$ denotes the code, and $\ell(\cdot) $ denotes the length function. In this case, for a discrete memoryless source $Y^n$, Campbell \cite{Campbell1965} shows that the minimum per-letter cost asymptotically achievable by prefix codes is given by the \emph{R\'enyi entropy} $H_\frac{1}{1+\lambda}(Y)$. Notice that $L_\lambda(Y^n)$ captures the notion of risk-aversion through the parameter $\lambda$ since
\begin{align}
	L_\lambda(Y^n) &\xrightarrow{\lambda \to 0}  \bbE\left[ \ell(\sff(Y^n)) \right]\text{,} \\
	L_\lambda(Y^n) &\xrightarrow{\lambda \to \infty} \max_{y^n \in \cY^n} \ell(\sff(y^n))\text{.}
\end{align}
A natural way to introduce risk-aversion in universal source coding is to use Campbell's formulation and characterize the penalty for the mismatch akin to \eqref{eqn:redundancy}. Indeed, about forty years after Campbell's work, Sundaresan \cite[Theorem 8]{Sundaresan2007} shows that if one uses $L_\lambda(Y^n)$ as the cost function, the penalty paid for universality can be written as\footnote{Campbell's and Sundaresan's results are still valid when $\lambda \in (-1,0)$. However, such a formulation corresponds to a \emph{risk-seeking} scheme, which falls outside the philosophy espoused in this paper.}
\begin{align}
	\frac{1}{n} D_{1+\lambda}(\widetilde{P}_{Y^n|V=\boldtheta}^{\frac{1}{1+\lambda} }\| \widetilde{Q}_{Y^n}^{\frac{1}{1+\lambda} } )+o(1)\text{,}
\end{align}
where $D_{1+\lambda}(P\|Q)$ denotes the \emph{R\'enyi divergence of order} $1+\lambda$, which is defined in \eqref{eqn:renyi divergence definition}, and $\widetilde{P}_Y^\alpha$ denotes the \emph{scaled distribution} of $P_Y$:
\begin{align}
	\widetilde{P}_Y^\alpha(y)=\frac{P_Y^\alpha(y)}{\sum_{b\in \cY }P_Y^\alpha (b)}\text{.} \label{eqn:scaled-distribution}
\end{align}
The distance measure
\begin{align}
	S_\alpha(P \| Q)=D_\alpha (\widetilde{P}^{ \frac{1}{\alpha}}\|\widetilde{Q}^{\frac{1}{\alpha}})
\end{align}
is known as the \emph{Sundaresan divergence} of order $\alpha$ between $P$ and $Q$. Following \cite{Sundaresan2007}, the relevant measure of redundancy for universal lossless compression under Campbell's performance criterion is 
\begin{align}
	R_\lambda(n)=\inf_{Q_{Y^n}} \sup_{\boldtheta} 	S_{1+\lambda}(P_{Y^n|V=\boldtheta} \| Q_{Y^n} )\text{.} \label{eqn:minimax renyi redundancy}
\end{align}
The conventional minimax redundancy in \eqref{eqn:minimax redundancy definition} corresponds to $R_0(n)$ while the minimax regret in \eqref{eqn:minimax regret definition} corresponds to $R_\infty(n)$. Although, in general, $ S_{\alpha} ( P \| Q) \neq D_{\alpha} ( P \|Q) $, we are able to establish a pleasing analog to the classical redundancy results such as \eqref{eqn:minimax redundancy definition}, \eqref{eqn:redundancy-capacity theorem} and \eqref{eqn:minimax regret definition}, \eqref{eqn:redundancy-capacity for Regret}:
\begin{align}
R_\lambda(n)&=\inf_{Q_{Y^n}} \sup_{\boldtheta  } 	D_{1+\lambda}(P_{Y^n|V=\boldtheta}  \| Q_{Y^n} ) \label{aaaa1} \\
&=\sup_{P_V} I_{1+\lambda} (P_V, P_{Y^n|V} )\text{,} \label{aaaa2} 
\end{align}	
where in \eqref{aaaa2}
\begin{align}
I_{1+\lambda}(P_X, P_{Y|X}) = \inf_{Q_Y} D_{1+\lambda} (P_{Y|X}P_X \| Q_Y P_X)
\end{align}
is the \emph{$\alpha$-mutual information of order} $1+\lambda$ between $X$ and $Y$ with $(X,Y)\sim P_XP_{Y|X}$, see \cite{sibson1969information}, \cite{Verdu2015}. Note that \eqref{aaaa1} is analogous to \eqref{eqn:minimax redundancy definition} with R\'enyi divergence replacing the relative entropy. Thus, we refer $R_\lambda(n)$ as the \emph{minimax R\'enyi redundancy}. Moreover, \eqref{aaaa2} generalizes the Redundacy-Capacity Theorem to $\alpha$-mutual information thereby finding another operational meaning for the maximal $\alpha$-mutual information beyond those that have been shown in the literature on error probability bounds for data transmission (e.g. \cite{Verdu2015}, \cite{Csiszar95}). Moreover, the $\alpha$-mutual information smoothly interpolates between two extremes, namely $I(P_V, P_{Y^n|V})$ in \eqref{eqn:redundancy-capacity theorem} and $I_\infty(P_V, P_{Y^n|V})$ in \eqref{eqn:redundancy-capacity for Regret}. Finally, \eqref{aaaa1} and \eqref{aaaa2}, coupled with Campbell's result~\cite{Campbell1965}, provide a pleasing relationship between R\'enyi entropy, R\'enyi	divergence and $\alpha$-mutual information in the context of universal lossless data compression.

The asymptotic behaviors of the minimax redundancy and minimax regret have also received considerable attention in the literature (e.g., \cite{XieBarron2000}, \cite{Shtarkov87}, \cite{Dramota04}, \cite{XieBarron1997, Davisson81, Gyorfi94, KriTro81, Rissanen84, Rissanen86}) since, in addition to compression, they are relevant in applications such as machine learning, finance, prediction, gambling, and so on. In particular, Xie and Barron in their key contributions \cite{XieBarron1997}, \cite{XieBarron2000}  show that 
\begin{align}
	R_n&=R_0(n) \\ 
	 &=\frac{k-1}{2}\log \frac{n}{2\pi \rme}+\log \frac{\opGamma^k(1/2)}{\opGamma(k/2)} + o(1)\text{,} \label{eqn:barron1} \\
	r_n&=R_\infty(n) \\
	&=\frac{k-1}{2}\log \frac{n}{2\pi}+\log \frac{\opGamma^k(1/2)}{\opGamma(k/2)} +o(1)\text{,} \label{eqn:barron2}
\end{align}
where $n$ and $k$ are the number of observations and the alphabet size, respectively, $\opGamma$ denotes the Gamma function, and $o(1)$ vanishes as $n\to \infty$.

While Merhav \cite[Theorem 1]{Merhav11} gives $R_\lambda(n) = \frac{k-1}{2}\log n +o(\log n) $, we quantify asymptotically the effect of the risk-aversion parameter $\lambda$ on the fundamental limit in universal source coding by providing a pleasing interpolation\footnote{In a fundamentally different setup, Hayashi \cite[Lemma 3]{Hayashi17} considers the counterpart of the Clarke and Barron \cite[Theorem 2.1]{ClarkeBarron} result replacing relative entropy with R\'enyi divergence.} between \eqref{eqn:barron1} and \eqref{eqn:barron2}: 
\begin{align}
 R_\lambda (n) = \frac{k-1}{2}\log \frac{n}{2\pi(1+\lambda)^{\frac 1\lambda} }+\log \frac{\opGamma^k(1/2)}{\opGamma(k/2)} 
 +o(1)\text{.}  
\label{aaab}
\end{align}

In the remainder of the paper, Section~\ref{sec:notation} sets the basic notation and definitions. Section~\ref{sec:main results} states the main results and gives the outlines of their proofs, which are contained in Section~\ref{sec:proofs}. In the Appendices, we prove several lemmas that are used in Section~\ref{sec:proofs}.

\section{Notation and Definitions}\label{sec:notation}
Let $\cY=\{1, 2, \hdots, k \}$ and denote the $(k-1)$-dimensional simplex of probability mass functions defined on $\cY$ by
\begin{align}
	\simplex^{k-1}=\left\{(\theta_1, \hdots, \theta_k)\in \bbR_+^k \colon \sum_{i=1}^{k}\theta_i = 1\right\}\text{.}
\end{align}
For each parameter $\boldtheta=(\theta_1, \hdots, \theta_k ) \in \simplex^{k-1}$, we define our observation model $P_{Y|V=\boldtheta}\colon \simplex^{k-1} \to \cY$ such that\footnote{As a special case, when $k=2$, we use the shorthand notation $P_{Y|V=\theta}$ instead of $P_{Y|V=(\theta,1-\theta)}$.}
\begin{align}
	P_{Y|V=\boldtheta}(i)=\theta_i\text{,}
\end{align}
and the independent identically distributed (i.i.d.) extension of this model $ P_{Y^n|V=\boldtheta}\colon  \simplex^{k-1}\to \cY^n $ such that
\begin{align}
P_{Y^n|V=\boldtheta}(y^n)&=\prod_{i=1}^{n}P_{Y|V=\boldtheta}(y_i) \\
&=\theta_1^{t_1} \cdots \theta_k^{t_k}\text{,} \label{eqn:model}
\end{align}
where
\begin{align}
	t_i=\sum_{j=1}^{n}1\{y_j=i\}\text{,} \label{eqn:type}
\end{align}
denotes the number of times $i\in \cY$ appears in the vector $y^n$, and therefore
\begin{align}
	\sum_{i=1}^{k}t_i=n\text{.}
\end{align}
It can be verified that the Fisher information matrix (in nats) of $P_{Y|V=\boldtheta}$ for the parameter vector $\boldtheta$ is\footnote{Note that the Fisher information matrix is $(k-1)\times (k-1)$ since there are $(k-1)$ free parameters in the model. Nevertheless, it is notationally convenient to denote the parameter vector $\boldtheta $ as if it were $k$-dimensional.}
\begin{align}
	\bfJ(\boldtheta, P_{Y|V} )&=\diag\left( \frac{1}{\theta_1}, \frac{1}{\theta_2}, \hdots, \frac{1}{ \theta_{k-1}} \right)+\frac{1}{\theta_k}\mathbf{1}_{(k-1)\times(k-1)}\text{,} \label{eqn:fisher_matrix}
\end{align}
where $\mathbf{1}_{l\times l} $ denotes an $l \times l $ matrix all of whose entries are equal to 1. The determinant of the Fisher information matrix in \eqref{eqn:fisher_matrix} satisfies
\begin{align}
	|\bfJ(\boldtheta, P_{Y|V})|=\frac{1}{\prod_{i=1}^{k} \theta_i }\text{.}
\end{align}
An important probability measure on $\simplex^{k-1} $ is \emph{Jeffreys' prior} \cite{Jeffreys46} defined as
\begin{align}
	P_{V}^\ast(\boldtheta)&=\frac{|\bfJ(\boldtheta, P_{Y|V})|^{1/2}}{\ds \int_{\simplex^{k-1} } |\bfJ(\boldxi, P_{Y|V})|^{1/2} \rmd \boldxi}\\
		&=\frac{\theta_1^{-1/2} \cdots \theta_k^{-1/2} }{\opD_k(1/2, \hdots, 1/2)} \text{,} \label{eqn:defn:jeffreys prior}
\end{align}
where $\opD_k(\alpha_1, \hdots, \alpha_k)$ denotes a special form of the Dirichlet integrals of type 1 which can be written in terms of the Gamma function:
\begin{align}
	\opD_k(\alpha_1, \hdots, \alpha_k)&= \int_{\simplex^{k-1} } \xi_1^{\alpha_1-1 } \cdots \xi_k^{\alpha_k-1 } \rmd \boldxi \\
		&= \frac{\opGamma(\alpha_1) \cdots \opGamma(\alpha_k ) }{\opGamma(\alpha_1+\cdots+\alpha_k ) }\text{.} \label{eqn:def:dirichlet integrals}
\end{align}
In particular,
\begin{align}
	\opD_k(1/2, \hdots, 1/2) &= \frac{\opGamma^k(1/2)}{\opGamma(k/2)} \\
	&= \begin{cases}
		\ds\frac{\pi^{k/2}}{(k/2-1)!}\text{,} & k \text{ is even,} \vspace{1em} \\
		\ds\frac{\pi^{(k-1)/2}}{\prod_{i=1}^{(k-1)/2}\left(i-\frac12\right)}\text{,} & k \text{ is odd.}
	\end{cases}
\end{align}

The source distribution we get by assuming Jeffreys' prior on the parameter space is referred as \emph{Jeffreys' mixture} which is denoted by\footnote{Whenever it is informative to explicitly show the dimensionality of the parameter space in the notation for Jeffreys' mixture, we do so by replacing $Q^\ast_{Y^n}$ with $Q^{\ast(k-1)}_{Y^n}$.}
\begin{align}
	Q^\ast_{Y^n}(y^n)&=\int_{\simplex^{k-1} } P_{Y^n|V=\boldtheta}(y^n)\rmd P_{V}^\ast(\boldtheta) \label{eqn:def:jeffreys mixture} \\
	&= \frac{\opD_k(t_1+1/2, \hdots, t_k+1/2 ) }{\opD_k(1/2, \hdots, 1/2)} \text{.}
\end{align}

For discrete probability measures $P$ and $Q$ on the set $\cY $ such that $Q$ dominates $P$, i.e., $P \ll Q $, \emph{R\'enyi divergence} of order\footnote{We are not concerned with R\'enyi divergences of order $\alpha\in (0,1) $. A more general definition can be found in \cite{Erven2014}.} $\alpha $ between $P$ and $Q$ is defined as
\begin{align}
	&D_{\alpha}(P\|Q) \nonumber \\
	&\ =
	\begin{cases}
	 D(P\|Q)\text{,}  &  \alpha =1 \\
	\frac{1}{\alpha-1}\log \bbE[ \exp( (\alpha-1) \imath_{P\|Q}(Y))]\text{,} &  \alpha \in (1,\infty) \\
		\ds\max_{b\in \cY } \imath_{P\|Q}(b)\text{,} & \alpha =\infty\text{,}
	\end{cases} \label{eqn:renyi divergence definition}
\end{align}
where $Y\sim P $. In particular, when $\alpha\in (1,\infty)$, R\'enyi divergence of order $\alpha$ between $P$ and $Q$ can be expressed as
\begin{align}
	D_{\alpha}(P\|Q)=\frac{1}{\alpha-1}\log\sum_{b\in \cY} P^\alpha(b) Q^{1-\alpha}(b)\text{.}
\end{align}
Given $(P_X, P_{Y|X})$, an analogous generalization can be made for mutual information resulting in the \emph{$\alpha$-mutual information}\footnote{The definition of $\alpha$-mutual information in \eqref{eqn:alpha-mutual-information definition} dates back to Sibson's information radius \cite{sibson1969information}. Although, it should be noted that Sibson's motivation in \cite{sibson1969information} is not the generalization of mutual information. See \cite{Verdu2015} for a more thorough discussion.} \cite{Verdu2015}:
\begin{align}
	& I_{\alpha}(P_X, P_{Y|X}) \nonumber \\ 
	&\ =
	\begin{cases}
	I(P_X, P_{Y|X})\text{,} &  \alpha=1 \\
	\ds\inf_{Q_Y} D_{\alpha} (P_{Y|X}P_X \| Q_Y P_X)\text{,} &  \alpha\in (1,\infty)   \\
	\ds\log\bbE\bigg[\esssup_{X} \exp\big(   \imath_{X; Y }(X;\bar{Y}) \big) \bigg]\text{,} &  \alpha=\infty\text{,}  
	\end{cases} \label{eqn:alpha-mutual-information definition} 
\end{align}
where $\bar{Y}\sim P_Y $, independent of $X\sim P_X $, and we have used the conventional notation for \emph{information density} $\imath_{X;Y}(x;y)=\imath_{P_{Y|X=x}\|P_{Y} }(y) $. As shown in Lemma~\ref{lem:explicit alpha mut info} in Appendix~\ref{appdx: Explicit alpha mutual info}, the infimum in \eqref{eqn:alpha-mutual-information definition} can be solved explicitly. 

In parallel with the standard usage for relative entropy, it is common to define the conditional R\'enyi divergence as
\begin{align}
	D_\alpha(P_{Y|X}\|Q_{Y|X}|P_X )=D_\alpha(P_X P_{Y|X} \| P_X Q_{Y|X} )\text{,}
\end{align}
therefore, the unconditional R\'enyi divergence in \eqref{eqn:alpha-mutual-information definition} can be written as $D_\alpha(P_{Y|X }\| Q_{Y}|P_X ) $.

\section{Statement of the Results}\label{sec:main results}

Theorem~\ref{thm:Generalized Redundancy-Capacity Theorem} states that under the minimax operation in \eqref{eqn:minimax renyi redundancy} the Sundaresan divergence can be replaced by the R\'enyi divergence. We further show that this minimax operation can be written as the maximization of the $\alpha$-mutual information, thus, providing a generalization to the Redundancy-Capacity Theorem in \eqref{eqn:redundancy-capacity theorem}. In Theorem~\ref{thm:Asymptotic Behavior of Minimax Renyi Redundancy}, we investigate the asymptotic behavior of the minimax R\'enyi redundancy between $P_{Y^n|V=\boldtheta}$ and $Q_{Y^n}$, and we find its precise asymptotic expansion, thereby quantifying the effect of the risk-aversion parameter $\lambda$.

\begin{thm}[\textit{\textsf{\small Generalized Redundancy-Capacity Theorem}}] \label{thm:Generalized Redundancy-Capacity Theorem} 
For any $\lambda \in (0,\infty) $, and positive integer $n$ 
	\begin{align}
R_\lambda(n)
&=\inf_{Q_{Y^n}} \sup_{\boldtheta\in \simplex^{k-1} }D_{1+\lambda}(P_{Y^n|V=\boldtheta}  \| Q_{Y^n} ) \label{eqn:thm:gen-red-cap:1}  \\
&=\sup_{P_V} I_{1+\lambda} (P_V, P_{Y^n|V})\text{.} \label{eqn:thm:gen-red-cap:2}
	\end{align}
\end{thm}

As we show in the proof in Section~\ref{sec:proofs}, \eqref{eqn:thm:gen-red-cap:1} is due to the fact that scaling a distribution is a one-to-one operation that preserves memorylessness while the minimax theorem for R\'enyi divergence \cite[Theorem 34]{Erven2014} is the gateway to showing the generalized redundancy-capacity theorem in \eqref{eqn:thm:gen-red-cap:2}. 

Although Theorem~\ref{thm:Generalized Redundancy-Capacity Theorem} holds in great generality, we illustrate its use in the simple example below. 
\begin{exa*}[\textit{\textsf{\small Z-Channel with $\frac12$ crossover probability}}] \normalfont
Consider the \emph{Z-channel} with $\frac12$ crossover probability, see, e.g., \cite[Problem 7.8]{CoverThomas}. In this case,
\begin{align}
	&\frac{\lambda}{1+\lambda} I_{1+\lambda}(P_V, P_{Y|V}) = \label{Z-channel formula for alpha mut inf} \\	
	&\qquad  \log \Bigg(\bigg(\frac{P_V(1)}{2^{1+\lambda}}\bigg)^{\frac{1}{1+\lambda}}+ \bigg(1-\frac{2^{1+\lambda} -1}{2^{1+\lambda}}P_V(1)\bigg)^{\frac1{1+\lambda}} \Bigg)  \nonumber
\end{align}
which is a concave function of $P_{V}(1)$ for every value of $\lambda\in (0,\infty) $, and is maximized when 
\begin{align}
	P_{V}(1) &=1- \frac{1-(2^{1+\lambda}-1)^{-\frac{1+\lambda}{\lambda}}}{1+(2^{1+\lambda}-1)^{-\frac{1}{\lambda}}} \text{.} \label{optimal choice}
\end{align}
After some elementary algebra, plugging \eqref{optimal choice} into \eqref{Z-channel formula for alpha mut inf} yields
\begin{align}
	\sup_{P_{V}} I_{1+\lambda}(P_V, P_{Y|V}) = \log\left(1+\left(2^{1+\lambda}-1\right)^{-\frac{1}{\lambda}}\right) \text{.} \label{exa:maximal alpha mutual}
\end{align}
Observe that as $\lambda \to 0$, the right side of \eqref{exa:maximal alpha mutual} converges to the capacity of the channel, namely, $\log \frac54$. 
On the other hand, to compute the minimax R\'enyi redundancy, note that 
\begin{align}
	D_{1+\lambda}(P_{Y|V=0}\|Q_Y) &= \log \frac{1}{Q_Y(0)} \text{,} \\
	D_{1+\lambda}(P_{Y|V=1}\|Q_Y) &= \frac1\lambda \log \left(\frac{2^{-(1+\lambda)}}{Q^\lambda_Y(0)}+\frac{2^{-(1+\lambda)}}{Q^\lambda_Y(1)}\right) \text{.}
\end{align}
Let $Q^\ast_Y$ be the distribution such that 
\begin{align}
	D_{1+\lambda}(P_{Y|V=0}\|Q^\ast_Y) = D_{1+\lambda}(P_{Y|V=1}\|Q^\ast_Y) \text{.}
\end{align}
Since
\begin{align}
	& \inf_{Q_Y}\sup_{\theta \in \{0,1\}} D_{1+\lambda}(P_{Y|V=\theta}\|Q_Y) \nonumber \\ 
	&\qquad = D_{1+\lambda}(P_{Y|V=0}\|Q^\ast_Y) \\ 
	&\qquad = \log\left(1+\left(2^{1+\lambda}-1\right)^{-\frac{1}{\lambda}}\right) \text{,} \label{exa:minmax renyi} 
\end{align}
through \eqref{exa:maximal alpha mutual} and \eqref{exa:minmax renyi}, as enforced by generalized redundancy-capacity theorem, we observe that the maximal $\alpha$-mutual information matches the minimax R\'enyi divergence. 
\end{exa*}

\begin{thm}[\textit{\textsf{\small Asymptotic Behavior of Minimax R\'enyi Redundancy}}] \label{thm:Asymptotic Behavior of Minimax Renyi Redundancy} For any $\lambda \in (0, \infty) $
\begin{align}
	&\lim_{n\to \infty} \left\{ 	R_\lambda (n)-\frac{k-1}{2}\log \frac{n}{2\pi} \right\} \nonumber \\
	&\qquad \qquad =\log \frac{\opGamma^k(1/2)}{\opGamma(k/2)}-\frac{k-1}{2\lambda}\log(1+\lambda) \text{.} \label{eqn:thm:asymp-minmax-renyi-red}
\end{align}		
\end{thm}
We prove Theorem~\ref{thm:Asymptotic Behavior of Minimax Renyi Redundancy} in Section~\ref{sec:proofs} by dividing it into two parts: converse and achievability. In both parts, Jeffreys' prior plays a significant role. However, it is known that Jeffreys' prior dramatically emphasizes the lower dimensional faces of the simplex. While this is not a problem in proving the converse bound, Jeffreys' prior achieves a suboptimal minimax value (see Lemma~\ref{lem:jefreys mixture is not minimax} in Appendix~\ref{appdx:jeffreys' mixture is not minimax}). Similar issues arise in finding the exact asymptotic constant in minimax redundancy \cite{XieBarron1997}, and in minimax regret \cite{XieBarron2000}. To overcome this problem, we modify Jeffreys' prior by placing masses near the faces of the simplex as in \cite{XieBarron1997}. Although this resolves the problem encountered in the minimax redundancy and minimax regret cases, the functional form of R\'enyi divergence becomes the second obstacle which forces us to show a uniform Laplace approximation thereby making the proof of achievability a much more involved task than that of the converse. For this reason, we start by presenting the achievability proof in the special case of binary alphabets, in which the notation is simplified considerably.

\section{Proofs } \label{sec:proofs}
\subsection{Proof of Theorem~\ref{thm:Generalized Redundancy-Capacity Theorem}}

To establish \eqref{eqn:thm:gen-red-cap:1}, for any $\lambda \in (0, \infty) $, define the bijection $f_\lambda \colon \simplex^{k-1}\to \simplex^{k-1}$ as
\begin{align}
f_\lambda(\theta_1, \hdots, \theta_k)=\frac1{\kappa_\lambda}\left( \theta_1^{\frac{1}{1+\lambda}}, \hdots, \theta_k^{\frac{1}{1+\lambda}} \right)\text{,}
\end{align}
where
\begin{align}
	\kappa_\lambda=\sum_{b\in \cY}P_{Y|V=\boldtheta}^\frac{1}{1+\lambda}(b)\text{.}
\end{align}
Then, for any $\boldtheta=(\theta_1, \hdots, \theta_k) \in \simplex^{k-1} $ and $y^n \in \cY^n $, the scaled version of the conditional distribution (see \eqref{eqn:scaled-distribution}) satisfies
\begin{align}
\widetilde{P}_{Y^n|V=\boldtheta}^{\frac{1}{1+\lambda}}(y^n)
&=
\prod_{i=1}^{n}\widetilde{P}_{Y|V=\boldtheta}^{\frac{1}{1+\lambda}}(y_i)\\
&=\prod_{i=1}^{n}P_{Y|V=f_\lambda(\boldtheta)}(y_i) \\
&=P_{Y^n|V=f_\lambda(\boldtheta)}(y^n)\text{.}
\end{align}
Therefore, for any given distribution $R_{Y^n}$ on $\cY^n$
\begin{align}
&\sup_{\boldtheta\in \simplex^{k-1}}D_{1+\lambda}(\widetilde{P}_{Y^n|V=\boldtheta}^{\frac{1}{1+\lambda}}\|R_{Y^n}) \nonumber \\
&\qquad =\sup_{\boldtheta\in \simplex^{k-1}}D_{1+\lambda}({P}_{Y^n|V= f_\lambda(\boldtheta)}\|R_{Y^n})  \\
&\qquad =\sup_{\boldtheta \in \simplex^{k-1}}D_{1+\lambda}({P}_{Y^n|V=\boldtheta}\|R_{Y^n})\text{.} \label{ekzz:P}
\end{align}
As a result of \eqref{ekzz:P},
\begin{align}
&\inf_{Q_{Y^n} }\sup_{\boldtheta\in \simplex^{k-1}} S_{1+\lambda}(P_{Y^n|V=\boldtheta}  \| Q_{Y^n}) \nonumber \\
&\qquad =
\inf_{Q_{Y^n} }\sup_{\boldtheta\in \simplex^{k-1}}D_{1+\lambda} \Big(\widetilde{P}_{Y^n|V=\boldtheta}^{\frac{1}{1+\lambda}}\|\widetilde{Q}_{Y^n}^{\frac{1}{1+\lambda}}\Big) \\ 
&\qquad = \inf_{Q_{Y^n} }\sup_{\boldtheta \in \simplex^{k-1}}D_{1+\lambda} \Big(P_{Y^n|V=\boldtheta} \| \widetilde{Q}_{Y^n}^{\frac{1}{1+\lambda}}\Big) \label{ekzz:from_P}\\
&\qquad =\inf_{\widetilde{Q}^{\frac{1}{1+\lambda}}_{Y^n} }\sup_{\boldtheta\in \simplex^{k-1}}D_{1+\lambda} \Big(P_{Y^n|V=\boldtheta} \| \widetilde{Q}_{Y^n}^{\frac{1}{1+\lambda}}\Big)\label{ekzz:fromf}\\
&\qquad = \inf_{Q_{Y^n} }\sup_{\boldtheta\in \simplex^{k-1}}D_{1+\lambda}\left(P_{Y^n|V=\boldtheta}\|Q_{Y^n}\right)\text{,} \label{ekzz:dumm}
\end{align}
where \eqref{ekzz:fromf} follows because every probability measure in $\simplex^{nk-1} $ is a scaled version of another probability measure in $\simplex^{nk-1} $.

In order to establish \eqref{eqn:thm:gen-red-cap:2}, note that
\begin{align}
&\inf_{Q_{Y^n}}\sup_{\boldtheta\in \simplex^{k-1}}D_{1+\lambda}\left(P_{Y^n|V=\boldtheta}\|Q_{Y^n}\right) \nonumber \\
&\qquad = \inf_{Q_{Y^n}}\sup_{P_V}\bbE \left[  D_{1+\lambda}\left(P_{Y^n|V}(\cdot | V) \|Q_{Y^n}\right) \right] \label{expl:notation} \\
&\qquad = \sup_{P_V}\inf_{Q_{Y^n}} \bbE \left[ D_{1+\lambda}\left(P_{Y^n|V}(\cdot| V) \|Q_{Y^n}\right) \right] \label{eqn:from_minimax_theorem} \\
&\qquad =\sup_{P_V}I_{1+\lambda} (P_V,P_{Y^n|V})\text{,} \label{aabb}
\end{align}
where the expectation in \eqref{expl:notation} is with respect to $V \sim P_V $, and \eqref{eqn:from_minimax_theorem} follows from \cite[Theorem 34]{Erven2014}, which holds when $\cY$ is finite. The right side of \eqref{eqn:from_minimax_theorem} is the maximal $\alpha$-mutual information of order\footnote{When both random variables are discrete, another generalization of mutual information, whose maximum also coincides with \eqref{aabb}, is put forward by Arimoto \cite{arimoto75information}. See \cite{Verdu2015} for further discussion of the various proposals of $\alpha$-mutual information.} $1+\lambda$ in the sense of Csisz\'ar, see \cite{Verdu2015} and \cite{Csiszar95}, which is known to equal maximal $I_{1+\lambda}$ (see \cite[Proposition 1]{Verdu2015}, and \cite[Theorem 5]{Csiszar95}) in the discrete parameter case. To see that \eqref{aabb} holds even when the parameter space is continuous, recall the definition of $\alpha$-mutual information, \eqref{eqn:alpha-mutual-information definition}, which can be written as
\begin{align}
&I_{1+\lambda} (P_V,P_{Y^n|V}) \nonumber  \\ 
&\quad =\inf_{Q_{Y^n} } \frac{1}{\lambda}\log \bbE\bigg[ \bbE \bigg[\exp\big( \lambda\,\imath_{P_{Y^n|V}\|Q_{Y^n}}(Y^n)\big)\big| V  \bigg]  \bigg]\text{,} \label{alternate form for alpha mutual inform}
\end{align}
and note that
\begin{align}
&\sup_{P_V}\inf_{Q_{Y^n}} \bbE \left[\lambda D_{1+\lambda}\left(P_{Y^n|V}(\cdot|V) \|Q_{Y^n}\right) \right] \nonumber \\
&\le \sup_{P_V} \inf_{Q_{Y^n} } \log \bbE\bigg[ \bbE \bigg[\exp\big( \lambda\, \imath_{P_{Y^n|V} \| Q_{Y^n} } (Y^n) \big)\big| V  \bigg]  \bigg] \label{baaa} \\
&\le \inf_{Q_{Y^n} } \log \hspace{-0.5mm} \bigg(\hspace{-0.8mm} \sup_{P_V} \bbE\bigg[ \bbE \bigg[\hspace{-0.8mm} \exp\big( \lambda\, \imath_{P_{Y^n|V} \| Q_{Y^n} } (Y^n) \big)\big| V  \bigg]  \bigg] \bigg) \label{baab} \\
&=\inf_{Q_{Y^n}} \sup_{\boldtheta \in \simplex^{k-1} } \lambda D_{1+\lambda}\left(P_{Y^n|V=\boldtheta} \| Q_{Y^n}\right)  \\
&=\inf_{Q_{Y^n}} \sup_{P_V} \bbE \left[\lambda D_{1+\lambda}\left(P_{Y^n|V}(\cdot|V) \|Q_{Y^n}\right) \right] \\
&=\sup_{P_V} \inf_{Q_{Y^n}} \bbE \left[\lambda D_{1+\lambda}\left(P_{Y^n|V}(\cdot|V) \|Q_{Y^n}\right) \right]\text{,} \label{baac}
\end{align}
where \eqref{baaa} follows from Jensen's inequality, \eqref{baab} follows from the fact that the maximin value is always less than or equal to the minimax value, and \eqref{baac} is again due to \cite[Theorem 34]{Erven2014}. \hfill \IEEEQED
\subsection{Proof of the Converse of Theorem~\ref{thm:Asymptotic Behavior of Minimax Renyi Redundancy} }\label{sec:converse}

This section is devoted to the proof of
\begin{align}
&\liminf_{n\to \infty} \left\{  R_\lambda(n) -\frac{k-1}{2}\log \frac{n}{2\pi} \right\} \nonumber \\ 
&\qquad \ge \log\frac{\opGamma^k(1/2)}{\opGamma(k/2)}-\frac{k-1}{2\lambda}\log(1+\lambda)\text{,} \label{eqn:renyi-redundancy-conv}
\end{align}
for any $\lambda \in (0, \infty) $. Define
\begin{align}
\cM_n & =\left\{\bolda=(a_1, \hdots, a_k)\in \bbZ^k_+ \colon \sum_{i=1}^{k}a_i = n \right\}\text{,} \label{eqn:def:set M} \\
\cM_{n, \delta} & =  \cM_n \cap \left\{ (a_1, \hdots, a_k)\in \bbZ^k_+ \colon \frac{n\delta}{k}\leq a_i \ \forall i \right\}\text{,} \label{eqn:def:M_delta}
\end{align}
for any $\delta \in (0,1)$. Let $\boldt = (t_1, \ldots, t_k)$, and consider the following
\begin{align}
&\frac{\lambda}{1+\lambda} R_\lambda(n)=\sup_{P_V}\frac{\lambda}{1+\lambda} I_{1+\lambda}(P_V,P_{Y^n|V}) \label{eqn:array1} \\
&= \sup_{P_V} \log \sum_{y^n\in\cY^n} \left( \int_{\simplex^{k-1}}P_{Y^n|V=\boldtheta}^{1+\lambda}(y^n)\rmd P_V(\boldtheta) \right)^{\frac{1}{1+\lambda}} \label{eqn:array2} \\
&\ge \log\hspace{-1mm} \sum_{ \boldt \in\cM_n}\hspace{-1mm} \binom{n}{\boldt }\hspace{-0.5mm}
\bigg(\hspace{-0.5mm} \int_{\simplex^{k-1}}(\theta_1^{t_1}\cdots\theta_k^{t_k})^{1+\lambda}\rmd P^*_{V}(\boldtheta)\bigg)^{\frac{1}{1+\lambda}}  \label{eqn:array3} \\
&\ge \log \hspace{-1.2mm}\sum_{ \boldt \in\cM_{n, \delta}}\hspace{-1.2mm} \binom{n}{\boldt} \hspace{-0.5mm}
\bigg(\hspace{-0.5mm}\int_{\simplex^{k-1}}(\theta_1^{t_1}\cdots\theta_k^{t_k})^{1+\lambda}\rmd P^*_{V}(\boldtheta)\hspace{-0.5mm} \bigg)^{\frac{1}{1+\lambda}} \text{,} \label{eqn:msubsetm}
\end{align}
where \eqref{eqn:array1} is due to Theorem~\ref{thm:Generalized Redundancy-Capacity Theorem}, \eqref{eqn:array2} follows from a more general result\cite[Theorem 1]{Verdu2015}, although, for the sake of completeness, its proof is included in Lemma \ref{lem:explicit alpha mut info} in Appendix \ref{appdx: Explicit alpha mutual info}, \eqref{eqn:array3} is due to the suboptimal choice of Jeffreys' prior, and \eqref{eqn:msubsetm} follows because $\cM_{n, \delta}\subset\cM_n$. 

Using Robbins' sharpening \cite{Robbins} of Stirling's approximation, one can show that
\begin{align}
&\frac{\rme^{nH(\widehat{P}_{y^n})}}{(2\pi)^{\frac{k-1}{2}}} \left(\frac{n}{\prod_{i=1}^k t_i}\right)^\frac12 
\frac{\rme^{\frac{1}{12(n+1)}}}{\prod_{i=1}^{k} \rme^{\frac{1}{12t_i}}} \nonumber \\
&\qquad \le \binom{n}{t_1 \cdots t_k } \label{eqn:lower bound on multinomial} \\
&\qquad \le \frac{\rme^{nH(\widehat{P}_{y^n})}}{(2\pi)^{\frac{k-1}{2}} } \left(\frac{n}{\prod_{i=1}^{k}t_i}\right)^\frac12
\frac{\rme^{\frac{1}{12n}}}{\prod_{i=1}^{k}\rme^{\frac{1}{12(t_i+1)}} } \text{.} \label{eqn:upper bound on multinomial}
\end{align}
where the entropy is in nats and $\widehat{P}_{y^n}$ denotes the empirical distribution of the vector $y^n$. Since $ \boldt \in \cM_{n, \delta}$, \eqref{eqn:lower bound on multinomial} particularizes to 
\begin{align}
	\binom{n}{t_1 \cdots t_k } \ge \frac{\rme^{nH(\widehat{P}_{y^n})}}{(2\pi)^{\frac{k-1}{2}}} \left(\frac{n}{\prod_{i=1}^k t_i}\right)^\frac12  
\frac{\rme^{\frac{1}{12(n+1)}}}{ \rme^{\frac{k^2}{12n\delta }}}\text{.} \label{eqn:low_bd_on_multinom}
\end{align}
With the aid of \eqref{eqn:defn:jeffreys prior} and \eqref{eqn:def:dirichlet integrals} we can express the integral in the right side of \eqref{eqn:msubsetm} as
\begin{align}
&\int_{\simplex^{k-1}}\left(\theta_1^{t_1}\cdots\theta_k^{t_k}\right)^{1+\lambda}\rmd P^*_{V}(\boldtheta) \nonumber \\
&\qquad = \frac{\prod_{i=1}^k\opGamma\left((1+\lambda)t_i+\frac 12\right)}{\opGamma\left((1+\lambda)n+\frac k2\right)}   \frac{1}{\opD_k\left(\frac12, \hdots, \frac12\right)} \text{.} \label{eqn:from_dirichletevaluate}
\end{align}
The gamma function generalization of Stirling's approximation (shown to be valid for positive real numbers by Whittaker and Watson \cite{Whittaker}) yields
\begin{align}
\opGamma(x)=\sqrt{2\pi}x^{x-1/2}\rme^{-x}(1+r), \quad x>0\text{,} \label{eqn:Stirling_Approximation_for_Gamma_Function}
\end{align}
where $|r|\leq \rme^{1/(12x)}-1$. In particular, for $i=1,\hdots,  k $,
\begin{align}
\begin{split}
\hspace{-0.5mm}\opGamma\Big((1+\lambda)t_i+\frac12\Big) &= \sqrt{2\pi}\Big((1+\lambda)t_i+\frac12\Big)^{(1+\lambda)t_i}  \\
 & \qquad  \times \rme^{-(1+\lambda)t_i - 1/2}(1+r_i)\text{,}
\end{split}  \label{gamma_stirling1}  \\
\begin{split}
\hspace{-0.5mm}\opGamma \hspace{-0.6mm} \Big(\hspace{-0.4mm} (1+\lambda)n+\frac k2\Big) &= \sqrt{2\pi}\Big((1+\lambda)n + \frac k2\Big)^{(1+\lambda)n+\frac{k-1}2} \\
&\qquad  \times \rme^{-(1+\lambda)n - k/2}(1+r_0)\text{,} 
\end{split} \label{gamma_stirling2}
\end{align}
where
\begin{align}
|r_i|&\leq \exp_\rme \left( \frac{1}{12(1+\lambda)t_i+6} \right) -1\text{,} \label{eqn:error:gamma stirling1}   \\
|r_0|&\leq \exp_\rme \left( \frac{1}{12(1+\lambda)n+6k}\right) -1 \label{eqn:error:gamma stirling2} \text{.}
\end{align}
It follows from \eqref{gamma_stirling1} and \eqref{gamma_stirling2} that
\begin{align}
&\frac{\prod_{i=1}^k\opGamma\left((1+\lambda)t_i+\frac 12\right)}{\opGamma\left((1+\lambda)n+\frac k2\right)}= \Big(\frac{2\pi}{n}\Big)^{\frac{k-1}{2}}  \frac{\rme^{-n(1+\lambda)H(\widehat{P}_{y ^n})}}{(1+\lambda)^{\frac{k-1}{2}}} \nonumber \\
&\qquad \qquad \times \frac{\prod_{i=1}^k\left(1+\frac{1}{2(1+\lambda)t_i}\right)^{(1+\lambda)t_i}(1+r_i)}{\left(1+\frac{k}{2(1+\lambda)n}\right)^{(1+\lambda)n+\frac{k-1}{2}} (1+r_0)}  \text{.} \label{eqn:gamma_approx}
\end{align}
Combining \eqref{eqn:from_dirichletevaluate} and \eqref{eqn:gamma_approx}, we can write
\begin{align}
&\int_{\simplex^{k-1}}\left(\theta_1^{t_1}\cdots\theta_k^{t_k}\right)^{1+\lambda}\rmd P^*_{V}(\boldtheta) \nonumber \\
&=\frac{(2\pi)^{\frac{k-1}{2}}}{\opD_k(\frac12, \hdots, \frac12)} 
\frac{\rme^{-n(1+\lambda)H(\widehat{P}_{y ^n})}}{(1+\lambda)^{\frac{k-1}{2}}n^{\frac{k-1}{2}}}  \label{eqn:integral_term_by_term} \\
&\qquad  \times \frac{\prod_{i=1}^k\left(1+\frac{1}{2(1+\lambda)t_i}\right)^{(1+\lambda)t_i}(1+r_i)}{\left(1+\frac{k}{2(1+\lambda)n}\right)^{(1+\lambda)n+\frac{k-1}{2}}(1+r_0)}  \nonumber \\
&\geq \frac{(2\pi)^{\frac{k-1}{2}}}{\opD_k(\frac12, \hdots, \frac12)} 
\frac{\rme^{-n(1+\lambda)H(\widehat{P}_{y ^n})}}{(1+\lambda)^{\frac{k-1}{2}}n^{\frac{k-1}{2}}}  \label{eqn:integral_final_lower_bound}\\ 
&\qquad \times \frac{ \left(1+\frac{k}{2(1+\lambda)n\delta}\right)^{(1+\lambda)n\delta }}{\left(1+\frac{k}{2(1+\lambda)n}\right)^{(1+\lambda)n+\frac{k-1}{2}}} 
 \frac{\left(2-\rme^{\frac{k}{12(1+\lambda)n\delta+6k}}\right)^k}{\rme^{\frac{1}{12(1+\lambda)n+6k}}}\text{,} \nonumber
\end{align}
where \eqref{eqn:integral_final_lower_bound} is due to the definition of $\cM_{n, \delta}$, \eqref{eqn:def:M_delta}, the fact that for any positive constant $c$, $(1+{c}/{x})^x$ is a monotone increasing function of $x$, and the fact that the error terms (see \eqref{eqn:error:gamma stirling1} and \eqref{eqn:error:gamma stirling2}) satisfy
\begin{align}
\frac{\prod_{i=1}^k(1+r_i)}{1+r_0}\geq \frac{\left(2-\rme^{\frac{k}{12(1+\lambda)n\delta+6k}}\right)^k}{\rme^{\frac{1}{12(1+\lambda)n+6k}}}\text{.} \label{eqn:stirling_error_lower_bnd}
\end{align}
Uniting the lower bounds in \eqref{eqn:msubsetm}, \eqref{eqn:low_bd_on_multinom} and \eqref{eqn:integral_final_lower_bound}, 
\begin{align}
&R_\lambda(n) -\frac{k-1}{2}\log\left(\frac{n}{2\pi}\right)
\geq-\frac{1}{\lambda}\log\opD_k(1/2, \hdots, 1/2)
\label{eqn:renyi-redundancy-conv-pf-final-1} \\
& 
-\frac{k-1}{2\lambda} \log(1+\lambda)+ \frac{1+\lambda}{\lambda}\log(\beta(n,\delta, k)\epsilon(n, \delta, k, \lambda))
\text{,}  \nonumber 
\end{align}
where
\begin{align}
\beta(n,\delta, k)&=\sum_{\boldt \in\cM_{n, \delta} }
\frac{1}{n^{k-1}} \frac{1}{\prod_{j=1}^k \left(\frac{t_j}{n}\right)^{1/2}}\text{,} \label{def:beta}\\
\epsilon(n, \delta, k, \lambda)&=
 \frac{\rme^{\frac{1}{12(n+1)}}}{ \rme^{\frac{k^2}{12n\delta}}}
\frac{ \left(1+\frac{k}{2(1+\lambda)n\delta}\right)^{n\delta }}{ \left(1+\frac{k}{2(1+\lambda)n}\right)^{n+\frac{k-1}{2(1+\lambda)}}} \label{def:epsilon} \\
&\qquad \times \left(\frac{\left(2-\rme^{\frac{k}{12(1+\lambda)n\delta+6k}}\right)^k}{\rme^{\frac{1}{12(1+\lambda)n+6k}}}\right)^{\frac{1}{1+\lambda}}\text{.} \nonumber
\end{align}
Notice that
\begin{align}
\lim_{n\to \infty}\epsilon(n, \delta, k, \lambda)&=1 \text{, for any }\delta\in(0,1)\text{,} \label{eqn:limit_epsilon1}\\
\lim_{\delta\to 0}\lim_{n\to \infty} \beta(n,\delta, k)&=\int_{ \simplex^{k-1}}\tau_1^{-1/2} \hdots \tau_k^{-1/2}\rmd \mathbold{\tau} \label{eqn:limit_beta1}\\
&=\opD_k(1/2, \hdots, 1/2)\text{,}
\label{eqn:limit_beta2}
\end{align}
where \eqref{eqn:limit_epsilon1} follows after noticing that each factor of $\epsilon(n,\delta, k)$ goes to 1, and \eqref{eqn:limit_beta1} follows from the definition of the Riemann integral. Assembling \eqref{eqn:renyi-redundancy-conv-pf-final-1}, \eqref{eqn:limit_epsilon1} and \eqref{eqn:limit_beta2}, we obtain the desired bound in \eqref{eqn:renyi-redundancy-conv}. \hfill \IEEEQED
\subsection{Proof of the achievability of Theorem~\ref{thm:Asymptotic Behavior of Minimax Renyi Redundancy} when $k=2$} \label{sec:subsec:ach-k=2}

In this section, we prove $\le $ in \eqref{eqn:thm:asymp-minmax-renyi-red} when $k=2$, i.e., 
\begin{align}
&\limsup_{n\to \infty} \left\{R_\lambda(n) -\frac{1}{2}\log \frac{n}{2\pi}\right\} \nonumber \\
&\qquad \le \log\frac{\opGamma^2(1/2)}{\opGamma(1)}-\frac{1}{2\lambda}\log(1+\lambda) \\ 
&\qquad =\log \pi -\frac{1}{2\lambda}\log(1+\lambda)  \text{.}
\label{eqn:ach-k=2-1}
\end{align}
To that end, we modify Jeffreys' prior by placing masses near the vertices of the simplex, i.e., $\simplex^1$, which, in turn, enables us to show that when the parameter\footnote{Since $k=2$, we have $\boldtheta=(\theta, 1-\theta)$. To simplify the discussion, we prefer the shorthand notation $\theta$ rather than $\boldtheta$.}  $\theta$  takes values near the vertices of the simplex the value of the minimax R\'enyi redundancy grows strictly slower than $\tfrac12 \log n + O(1)$. Thus, we focus on values of $\theta$ that are not close to the vertices of the simplex, thereby enabling us to argue that the minimax R\'enyi redundancy behaves as in \eqref{eqn:ach-k=2-1}.   

Inspired by Xie and Barron's \cite{XieBarron1997} modified Jeffreys' prior, for $\epsilon \in (0,1) $ and $c\in (0, 1/(2\log \rme)) $, consider the prior 
\begin{align}
&P_V^\epsilon(\theta)=(1-\epsilon) P^\ast _{V}(\theta)  \label{eqn:Mod_Jeff_1_dim} \\ 
&\qquad +\frac{\epsilon}{2} 1\left\{\theta=\frac{c\log n}{n}\right\}+\frac{\epsilon}{2} 1\left\{\theta=1- \frac{c\log n}{n}\right\}\text{,}  \nonumber
\end{align}
which differs from the one in \cite{XieBarron1997} in the location of the point masses. Because of the modification on Jeffreys' prior, the corresponding $Y^n$ marginal changes from $Q_{Y^n}^\ast $ in \eqref{eqn:def:jeffreys mixture} to
\begin{align}
Q^\epsilon _{Y^n}=(1-\epsilon) Q^\ast _{Y^n}+\frac{\epsilon}{2} P_{Y^n | V=\frac{c\log n}{n}}+\frac{\epsilon}{2} P_{Y^n | V=1-\frac{c\log n}{n}}\text{.} \label{eqn:mod_jeff_mix}
\end{align}
In view of Theorem~\ref{thm:Generalized Redundancy-Capacity Theorem},
\begin{align}
R_\lambda(n)
&\le
\sup_{\theta \in \left[ 0,1\right] }D_{1+\lambda }\left( P_{Y^n|V=\theta} \| Q_{Y^n}^{\epsilon}\right) \label{eqn:before three_sup} \\
&=
\max \left\{\Xi_1(n,\lambda, \epsilon), \Xi_2(n,\lambda, \epsilon), \Xi_3(n,\lambda, \epsilon)   \right\}\text{,} \label{eqn:three_sup}
\end{align}
where
\begin{align}   
\Xi_1(n,\lambda, \epsilon)&=
\sup _{\theta \in \left[ 0,\frac{c\log n}{n}\right] }  D_{1+\lambda } ( P_{Y^n|V=\theta} \| Q_{Y^n}^{\epsilon})\text{,} \label{eqn:def:Xi1} \\
\Xi_2(n,\lambda, \epsilon)
&=\hspace{-0.9mm} \sup _{\theta \in \left[\frac{c\log n}{n} ,1-\frac{c\log n}{n}\right] } \hspace{-1.1mm} D_{1+\lambda }( P_{Y^n|V=\theta} \| Q_{Y^n}^{\epsilon})\text{,} \label{eqn:def:Xi2} \\
\Xi_3(n,\lambda, \epsilon)&=
\sup_{\theta \in \left[ 1-\frac{c\log n}{n}, 1\right] }D_{1+\lambda }( P_{Y^n|V=\theta} \| Q_{Y^n}^{\epsilon})\text{.} \label{eqn:def:Xi3}
\end{align}
The following result shows that the first and the third supremizations in the right side of \eqref{eqn:three_sup} are both dominated by $\tfrac12 \log n + O(1)$. 
\begin{prop}\label{prop:achievability k=2 theta in face of simplex} 
If $c\in (0, 1/(2\log \rme) ) $, then
\begin{align} 
\max\left\{ 
\Xi_1(n,\lambda, \epsilon), \Xi_3(n,\lambda, \epsilon)
 \right\}
& \le
\log \frac{2}{\epsilon} +\frac{c(\log \rme)\log n }{1-\frac{c\log n}{n}}\text{.}
\end{align} 
\end{prop}
\begin{IEEEproof}
Assume that $\theta \in \left[ 0,\frac{c\log n}{n}\right]$. We have 
\begin{align}
&D_{1+\lambda }\left( P_{Y^n|V=\theta} \| Q_{Y^n}^{\epsilon}\right) \nonumber \\
&\qquad \le 
\log \frac{2}{\epsilon} +nD_{1+\lambda }\left( P_{Y|V=\theta}\|P_{Y | V=\frac{c\log n}{n}} \right)
 \label{bbaa}\\
&\qquad  \le
\log \frac{2}{\epsilon} +nD_{1+\lambda }\left( P_{Y|V=0}\|P_{Y | V=\frac{c\log n}{n}} \right) \label{bbab} \\
&\qquad =
\log \frac{2}{\epsilon} -n\log \left( 1-\tfrac{c\log n}{n} \right) \\
&\qquad  \le
\log \frac{2}{\epsilon} +\frac{c\log \rme}{1-\frac{c\log n}{n}} \log n \text{,} \label{bbac}
\end{align}
where \eqref{bbaa} follows from \eqref{eqn:mod_jeff_mix}, \eqref{bbab} follows because R\'enyi divergence is monotone decreasing in $\theta$ (see Lemma~\ref{lem:renyi divergence monotone decreasing} in Appendix~\ref{appdx:monotonicity of binary renyi divergence}) and \eqref{bbac} follows because, for $x<1 $, 
\begin{align}
\log\left( \frac{1}{1-x} \right) \le \frac{x}{1-x} \log \rme\text{.}
\end{align}
Using a symmetrical argument, one can show that the upper bound in \eqref{bbac} still holds when $\theta \in \left[ 1-{c\log n}/{n}, 1\right]$.
\end{IEEEproof} 
It remains to investigate the behavior of the second supremization in the right side of \eqref{eqn:three_sup}. Let
\begin{align}
\Xi_{\ast}(n, \lambda) = \sup_{\theta \in \left[ \frac{c\log n}{n}, 1-\frac{c\log n}{n} \right]} D_{1+\lambda }( P_{Y^n|V=\theta} \| Q_{Y^n}^{\ast}) \text{,}
\end{align}
and note that
\begin{align}
\Xi_2(n,\lambda, \epsilon)
&\le  
\log \frac{1}{1-\epsilon}+\Xi_{\ast}(n, \lambda)  \text{,} \label{bcaa}
\end{align}
which follows from \eqref{eqn:mod_jeff_mix}. The following proposition gives an asymptotic upper bound on $\Xi_{\ast}(n, \lambda)$.
\begin{prop} \label{prop:achievability-k=2-hard-step}
Let $c \in (0, 1/(2\log \rme))$. For any $ \lambda \in (0,\infty) $,
\begin{align}
&\limsup_{n \to \infty} \Bigg\{ \Xi_{\ast}(n, \lambda)-\frac{1}{2}\log \frac{n}{2\pi} \Bigg\} \nonumber \\ 
&\qquad  \le \log \frac{\opGamma^2(1/2)}{\opGamma(1)}-\frac{1}{2\lambda}\log (1+\lambda)\text{.}
\end{align}
\end{prop}
\begin{IEEEproof}
Let $\theta_1=\theta$ and $\theta_2=1-\theta$. Without loss of generality, we may assume that $\theta_1\le 1/2$, otherwise we may interchange the roles of $\theta_1$ and $\theta_2$ together with the roles of $t_1$ and $t_2=n-t_1$ below. Note that
\begin{align}
&D_{1+\lambda}(P_{Y^n|V=\theta_1}\| Q^{\ast}_{Y^n})  \nonumber \\
&\qquad = \frac{1}{\lambda} \log (\mfrakV(\lambda, \theta_1, n) +\mfrakW(\lambda, \theta_1, n))\text{,} \label{eqn:term_in_square_bracket}
\end{align}
where
\begin{align}
&\mfrakV(\lambda, \theta_1, n) \nonumber \\
&\quad = \left( \theta_1^{n(1+\lambda)}+  \theta_2^{n(1+\lambda)} \right)\left(  \frac{\opD_{2}(\frac12, \frac12)}{\opD_2(\frac12, n+\frac12)} \right)^\lambda\text{,} \label{eqn:def:frakV-k=2} \\
&\mfrakW(\lambda, \theta_1, n) \nonumber \\
&\quad =\sum_{t_1=1}^{n-1}\hspace{-0.5mm} \binom{n}{t_1} \hspace{-0.5mm} \left( \theta_1^{t_1} \theta_2^{t_2} \right)^{1+\lambda} \hspace{-0.5mm}\bigg(\frac{\opD_{2}(\frac12, \frac12)}{\opD_2(t_1+\frac12, t_2+\frac12)} \bigg)^\lambda\text{.} \label{eqn:def:frakW-k=2}
\end{align}
Thanks to Lemma~\ref{lem:edge cases of ti} in Appendix~\ref{appdx:edge cases of t_i}, we know that for all sufficiently large $n$ satisfying
\begin{align}
	\frac{k\ln n}{2n}<1\text{,}
\end{align}
we have
\begin{align}
\mfrakV(\lambda, \theta_1, n) \le  2C_2^\lambda(2) C_3^\lambda(2)n^{-(1+\lambda)c\log \rme }n^{\frac{\lambda}{2}}\text{,} \label{k=2_upp_bound_1}
\end{align}
where the explicit expressions for $C_2(k)$ and $C_3(k)$ are given in \eqref{eqn:C_2} and \eqref{eqn:C_3}, respectively. Hence, we may now focus attention on $\mfrakW(\lambda, \theta_1, n) $. Note that
\begin{align}
 \binom{n}{t_1} &\le \left(\frac{n}{2\pi t_1t_2}\right)^\frac12 
\exp_\rme\left(nh\left(\tfrac{t_1}{n}\right)+\tfrac{1}{12n}\right)  \label{bdaa} \text{,} \\
 \theta_1^{t_1} \theta_2^{t_2}  &=\exp_\rme \left(-n\left[ d\left(\tfrac{t_1}{n} \|\theta_1\right)+h\left(\tfrac{t_1}{n}\right) \right] \right) \text{,} \label{bdab} 
\end{align}
where $h \colon [0,1]\to [0,1]$ and $d(\cdot \| \cdot ) \colon [0,1]\times [0,1] \to [0,\infty]$ denote the binary entropy and the binary relative entropy functions in nats, respectively and the bound in \eqref{bdaa} follows from Stirling's approximation, see \eqref{eqn:upper bound on multinomial}. Note also that
\begin{align}
&\frac{1}{\opD_2(t_1+\frac12, t_2+\frac12)} = \frac{\opGamma(n+1) }{\opGamma(t_1+\frac12) \opGamma(t_2+\frac12) } \\
&\qquad \le  \left( \frac{n}{2\pi}  \right)^\frac{1}{2}  \frac{\rme^{n  h\left( \frac{t_1}{n} \right)}\left( 1+\frac{1}{n} \right)^{n+\frac{1}{2}}\rme^{\frac{1}{12(n+1)}}}{\prod_{i=1}^2 \left( 1+\frac{1}{2t_i} \right)^{t_i}\left( 2-\rme^{\frac{1}{12t_i+6}} \right) } \text{,} \label{bdbc}
\end{align}
 where \eqref{bdbc} also follows from an application of Stirling's approximation, see \eqref{eqn:Stirling_Approximation_for_Gamma_Function}.

By substituting \eqref{bdaa}, \eqref{bdab}, and \eqref{bdbc} into the right side of \eqref{eqn:def:frakW-k=2}, we get
\begin{align}
\mfrakW(\lambda, \theta_1, n) \le 
\left( \frac{n}{2\pi}  \right)^\frac{\lambda}{2} \opD_2^\lambda\left(\tfrac12, \tfrac12\right)\mfrakS (\lambda, \theta_1, n) \text{,} \label{k=2_upper_bound_2}
\end{align}
where
\begin{align}
\mfrakS(\lambda, \theta_1, n)=
&\sum_{t_1=1}^{n-1} \Big(\frac{n}{2\pi t_1t_2} \Big)^\frac12 \exp_\rme(-n(1+\lambda) d(\tfrac{t_1}{n}\|\theta_1)) \nonumber \\ 
&\qquad \times K(\lambda, n, t_1)\text{,} \label{eqn:def:frakS-k=2}
\end{align}
and
\begin{align}
K(\lambda, n, t_1)=\rme^\frac{1}{12n} \Bigg( \frac{( 1+\frac{1}{n} )^{n+\frac{1}{2}}\rme^{\frac{1}{12(n+1)}}}{\prod_{i=1}^2 ( 1+\frac{1}{2t_i})^{t_i}( 2-\rme^{\frac{1}{12t_i+6}})}    \Bigg)^\lambda\text{.} \label{eqn:def:K-when_k=2}
\end{align}
Note that we can find an asymptotically suboptimal upper bound on $\mfrakS (\lambda, \theta_1, n)$ that depends only on $\lambda $ by invoking Lemma~\ref{lem:unif_upp_bd_on_K} in Appendix~\ref{appdx:bounds on K}, which shows a non-asymptotic uniform upper bound on $K(\lambda, n, t_1)$, and then by invoking Lemma~\ref{lem:uniform bound on the sum} in Appendix~\ref{appdx:uniform upper bound on the sum}, which shows a non-asymptotic uniform upper bound on
\begin{align}
\mfrakT(\lambda, \theta_1, n)=
\sum_{t_1=1}^{n-1}  \Big(\frac{n}{2\pi t_1t_2} \Big)^\frac12 \exp_\rme(-n(1+\lambda) d(\tfrac{t_1}{n}\|\theta_1))\text{.} \label{eqn:def:mfrakT-k=2}
\end{align}
Finding the optimal upper bound, on the other hand, requires a uniform Laplace approximation on $\mfrakS(\lambda, \theta_1, n)$, which is introduced next. First, given $\delta \in (0,1)$, split $\mfrakS(\lambda, \theta_1, n) $ as
\begin{align}
\mfrakS(\lambda, \theta_1, n)&= S_1(\lambda, \theta_1, n, \delta) +S_2(\lambda, \theta_1, n, \delta) \label{eqn:key_splitting} \\ 
&\qquad +S_3(\lambda, \theta_1, n, \delta)\text{,}  \nonumber
\end{align}
where  
\begin{align}
S_1(\lambda, \theta_1, n, \delta)&= \sum_{t_1=1}^{\lfloor n(1-\delta)\theta_1 \rfloor}   \Big(\frac{n}{2\pi t_1t_2} \Big)^\frac12 \label{eqn:k=2_sum_S1} \\ & \nonumber
\quad \times \exp_\rme\left(-n(1+\lambda) d\left(\tfrac{t_1}{n}\|\theta_1\right)\right) K(\lambda, n, t_1)\text{,}  \\ 
S_2(\lambda, \theta_1, n, \delta)&= \sum_{t_1=\lceil n(1-\delta)\theta_1 \rceil }^{\lfloor n(1+\delta)\theta_1 \rfloor}  \Big(\frac{n}{2\pi t_1t_2} \Big)^\frac12  \label{eqn:k=2_sum_S2} \\ & \nonumber
\quad \times\exp_\rme\left(-n(1+\lambda) d\left(\tfrac{t_1}{n}\|\theta_1\right)\right)  K(\lambda, n, t_1)\text{,} \\
S_3(\lambda, \theta_1, n, \delta)&= \sum_{t_1=\lceil n(1+\delta)\theta_1 \rceil }^{n-1}   \Big(\frac{n}{2\pi t_1t_2} \Big)^\frac12 \label{eqn:k=2_sum_S3}\\ & \nonumber
\quad \times \exp_\rme\left(-n(1+\lambda) d\left(\tfrac{t_1}{n}\|\theta_1\right)\right) K(\lambda, n, t_1)\text{.} 
\end{align}
In Lemmas~\ref{lem:k=2_sum_S1_asympt}, \ref{lem:k=2_sum_S2_asympt} and \ref{lem:k=2_sum_S3_asympt} in Appendix~\ref{appdx:achievebility_k=2}, we show each of the following properties: 
\begin{align}
&\lim_{n \to \infty} \sup_{\theta_1\in [ \frac{c\log n}{n}, \frac{1}{2} ]} S_1(\lambda, \theta_1, n, \delta) =0 \quad \forall \delta\in (0,1) \text{,} \label{eqn:reslt:S1}  \\
&\lim_{\delta\to 0} \limsup_{n \to \infty} \sup_{\theta_1\in [ \frac{c\log n}{n}, \frac{1}{2} ]} S_2(\lambda, \theta_1, n, \delta) \le (1+\lambda)^{-\frac12 }\text{,} \\
&\lim_{n \to \infty} \sup_{\theta_1\in [ \frac{c\log n}{n}, \frac{1}{2} ]} S_3(\lambda, \theta_1, n, \delta) =0 \quad \forall \delta \in (0,1)\text{.} \label{eqn:reslt:S3}
\end{align}
Since the left side of \eqref{eqn:key_splitting} does not depend on $\delta $, \eqref{eqn:reslt:S1}--\eqref{eqn:reslt:S3} imply, by letting $\delta \to 0 $, that
\begin{align}
 \limsup_{n \to \infty} \sup_{\theta_1\in \left[ \frac{c\log n}{n}, \frac{1}{2} \right]} \mfrakS(\lambda, \theta_1, n) \le (1+\lambda)^{-\frac12}\text{.} \label{eqn:lim_sup_the_sup_of_sum}
\end{align}
Finally, it follows from \eqref{eqn:term_in_square_bracket}, \eqref{k=2_upp_bound_1}, \eqref{k=2_upper_bound_2}, and \eqref{eqn:lim_sup_the_sup_of_sum} that
\begin{align}
&\limsup_{n \to \infty} \left\{  \sup_{\theta_1 \in [ \frac{c\log n}{n}, \frac{1}{2} ]} D_{1+\lambda}(P_{Y^n|V=\theta_1}\| Q^{\ast}_{Y^n}) -\frac{1}{2}\log \frac{n}{2\pi} \right\} \nonumber \\ 
&\qquad \qquad \le \log \frac{\opGamma^2(1/2)}{\opGamma(1)}-\frac{1}{2\lambda}\log(1+\lambda)\text{.} \label{aooz:last_1}
\end{align}
Since $\theta_1+\theta_2=1$, it also follows that
\begin{align}
&\limsup_{n \to \infty} \left\{ \sup_{\theta_2 \in [ \frac{1}{2}, 1-\frac{c\log n}{n}]} \hspace{-2mm} D_{1+\lambda}(P_{Y^n|V=\theta_2}\| Q^{\ast}_{Y^n}) -\frac{1}{2}\log \frac{n}{2\pi}\right\} \nonumber \\
&\qquad \qquad \le \log \frac{\opGamma^2(1/2)}{\opGamma(1)}-\frac{1}{2\lambda}\log(1+\lambda)\text{.} \label{aooz:last_2}
\end{align}
Combining \eqref{aooz:last_1} and \eqref{aooz:last_2} gives us the promised result of Proposition~\ref{prop:achievability-k=2-hard-step}. 
\end{IEEEproof}
Invoking Proposition~\ref{prop:achievability k=2 theta in face of simplex}, we see that the functions in \eqref{eqn:def:Xi1} and \eqref{eqn:def:Xi3} can be bounded by
\begin{align}
\Xi_1(n,\lambda, \epsilon) -\frac{1}{2}\log \frac{n}{2\pi}
& \le \left(\frac{c\log \rme }{1-\frac{c\log n}{n}}-\frac{1}{2} \right)\log n  \label{aooy:aha_last1}  \\ 
&\qquad \quad +\frac{1}{2}\log ( 2\pi) + \log \frac{2}{\epsilon}\text{,} \nonumber \\
\Xi_3(n,\lambda, \epsilon) -\frac{1}{2}\log \frac{n}{2\pi} 
& \le \left(\frac{c\log\rme}{1-\frac{c\log n}{n}}-\frac{1}{2}\right)\log n\label{aooy:aha_last2} \\
&\qquad \quad +\frac{1}{2}\log(2\pi)+\log\frac{2}{\epsilon}\text{,}  \nonumber
\end{align} 
while thanks to \eqref{bcaa} and Proposition~\ref{prop:achievability-k=2-hard-step}, it follows that
\begin{align}
&\limsup_{n\to \infty} \left\{  \Xi_2(n,\lambda, \epsilon) - \frac{1}{2}\log \frac{n}{2\pi} \right\}
 \nonumber \\
&\qquad \le \log \frac{\opGamma^2(1/2)}{\opGamma(1)} -\frac{1}{2\lambda}\log(1+\lambda) +\log \frac{1}{1-\epsilon}  \text{.} \label{dd:finishes-achiv-k=2}
\end{align}
Since $c \in (0, 1/(2\log \rme))$, we see that the right side of \eqref{dd:finishes-achiv-k=2} asymptotically dominates the right sides of \eqref{aooy:aha_last1} and \eqref{aooy:aha_last2}. Due to \eqref{eqn:three_sup}, and \eqref{aooy:aha_last1}--\eqref{dd:finishes-achiv-k=2}, the desired result in \eqref{eqn:ach-k=2-1} follows by choosing an arbitrarily small $\epsilon $ in \eqref{eqn:Mod_Jeff_1_dim}.
\hfill\IEEEQED

\subsection{Proof of the achievability of Theorem~\ref{thm:Asymptotic Behavior of Minimax Renyi Redundancy} when $k>2$} \label{sec:subsec:ach-gen}
In this section, we prove $\le $ in \eqref{eqn:thm:asymp-minmax-renyi-red} when $k>2$, i.e.,
\begin{align}
&\limsup_{n\to \infty} \left\{ R_\lambda(n)  -\frac{k-1}{2}\log \frac{n}{2\pi} \right\} \nonumber \\ 
&\qquad \le \log \frac{\opGamma^k(1/2) }{\opGamma(k/2) }-\frac{k-1}{2\lambda} \log (1+\lambda) \text{.} \label{eqn:achievability-to-show-general}
\end{align}
To do so, we once again modify Jeffreys' prior as in the previous section by placing masses near the lower dimensional faces of the simplex, i.e., $\simplex^{k-1} $, which, in turn, enables us to show that when the parameter vector $\boldtheta$ takes values near the faces of the simplex, the value of the minimax R\'enyi redundancy grows strictly slower than $ \frac{k-1}{2} \log n +O(1) $. Hence, by focusing on the parameter values that are not close to the faces of the simplex, we show that the minimax R\'enyi redundancy behaves as in \eqref{eqn:achievability-to-show-general}.

Following the idea in \cite{XieBarron1997}, let $c\in (0, 1/(2\log \rme))$ and, for $i=1,\hdots, k $, define
\begin{align}
\cL_i=\left\{ \boldtheta \colon \theta_i=\frac{c\log n}{n}\right\} \cap \simplex^{k-1}. 
\end{align}
Accordingly, we define the probability measure $\mu_i$ with respect to $\rmd_i\boldxi=\rmd\xi_1 \cdots \rmd\xi_{i-1} \rmd\xi_{i+1} \cdots \rmd\xi_k$, the Lebesgue measure on $\bbR^{k-2}$, as
\begin{align}
\mu _{i}\left( \boldtheta \right) =\dfrac {\theta^{-1/2} _{1}\cdots \theta^{-1/2} _{i-1}\theta^{-1/2} _{i+1}\cdots \theta^{-1/2} _{k}}{\ds\int _{\cL_i}\xi^{-1/2}_{1}\cdots \xi^{-1/2}_{i-1}\xi^{-1/2} _{i+1}\cdots \xi^{-1/2}_k \rmd_i\boldxi}\text{.}
\end{align}
Finally, for $\epsilon\in (0,1) $, we define the prior distribution $P^{\epsilon }_{V}$ on the probability simplex $\simplex^{k-1}$ as
\begin{align}
P^{\epsilon }_{V}=\dfrac {\epsilon }{k}\sum ^{k}_{i=1}\mu _{i}+\left( 1-\epsilon \right) P^{\ast }_{V}\text{,} \label{eqn:mod_jeff_prior_k-1}
\end{align}
where $P_V^\ast $ is Jeffreys' prior. Because of the modification on Jeffreys' prior in \eqref{eqn:mod_jeff_prior_k-1}, the corresponding $Y^n$ marginal changes from $Q_{Y^n}^\ast $ in \eqref{eqn:def:jeffreys mixture} to
\begin{align}
Q^{\epsilon }_{Y^n}\left( y^n\right)
&=
\dfrac {\epsilon }{k}\sum ^{k}_{i=1}M_{i}\left( y^{n}\right) +\left( 1-\epsilon \right) Q^{\ast }_{Y^n}\left( y^{n}\right)\text{,} \label{eqn:modified_jeffrey_k-1}
\end{align} 
where
\begin{align}
&M_{i}\left( y^{n}\right) = \int _{\cL_i}P_{Y^n | V=\boldtheta }\left( y^{n}\right) \mu_{i}\left( \boldtheta \right) \rmd_{i}\boldtheta \\
&=  \left(\frac{c\log n}{n}\right)^{t_i}\left(1-\frac{c\log n}{n}\right)^{n-t_i}\label{eqn:m_i}\\ 
&\ \times\dfrac {\opD_{k-1}\left( t_1+\frac12,\ldots ,t_{i-1}+\frac12,t_{i+1}+\frac12,\ldots ,t_{k}+\frac12\right) }{\opD_{k-1}\left( 1/2,\ldots ,1/2\right)}\text{.} \nonumber
\end{align}
Define, for $i=1,\hdots, k$,
\begin{align}
\cR_i&=\left\{ \boldtheta \colon \theta _{i}\in \left[0, \dfrac {c\log n}{n} \right]\right\}  \label{eqn:definition_of_R_i}\text{,} \\
\cR_0&=\simplex^{k-1} -\bigcup^{k}_{i=1}\cR_i\text{.} \label{eqn:definition_of_R}
\end{align}
Note that $\cR_0$ denotes the vectors none of whose coordinates are within close proximity of zero in the sense of \eqref{eqn:definition_of_R_i}. 

In view of Theorem~\ref{thm:Generalized Redundancy-Capacity Theorem},
\begin{align}
R_\lambda (n)&=
\inf_{Q_{Y^n}}\sup_{\boldtheta\in \simplex^{k-1}}D_{1+\lambda}(P_{Y^n|V=\boldtheta}\|Q_{Y^n}) \\
&\le
\sup _{\boldtheta \in \simplex^{k-1} }D_{1+\lambda }( P_{Y^n|V=\boldtheta} \| Q_{Y^n}^{\epsilon}) \\
&=
\max_{i\in \{0,1, \hdots, k \}}    
\sup _{\boldtheta \in \cR_i}  D_{1+\lambda }(P_{Y^n|V=\boldtheta} \| Q_{Y^n}^{\epsilon})\text{.}  \label{eqn:max_of_sups_k-1}
\end{align}
The following result shows that the supremizations over $\cR_i$ for $i=1, \hdots, k$ in \eqref{eqn:max_of_sups_k-1} are all dominated by $\tfrac{k-1}{2}\log n +~O(1)$.
\begin{prop}\label{prop:achievability-general-faces dont matter}
If $c\in (0, 1/(2\log \rme))$, then for each $i\in \{1, \hdots, k \}$
\begin{align}
&\sup_{\boldtheta \in \cR_i}D_{1+\lambda}(P_{Y^n|V=\boldtheta} \| Q_{Y^n}^{\epsilon}) \le  \\&
\quad \log \frac{k}{\epsilon} + \log C_1(k-1) +  \left(\frac{k-2}{2}+\frac{c\log \rme }{1-\frac{c\log n }{n}} \right) \log n\text{,} \nonumber
\end{align}
where the explicit value of $C_1(k)$ is given in \eqref{eqn:C_1(k)}.
\end{prop}
\begin{IEEEproof}
Thanks to the symmetry, it suffices to show the result for $i=1$. To that end, define $f\colon \simplex^{k-1}\to \simplex^{k-2} $ as
\begin{align}
	f(\boldtheta)=\left(\ds \frac{\theta_2}{1-\theta_1}, \cdots, \frac{\theta_k}{1-\theta_1} \right)\text{,}
\end{align}
and let $Q^{\ast(k-2)}_{Y^n}$ denote the Jeffreys' mixture when the underlying parameter space is the $(k-2)$-dimensional simplex. Further define
\begin{align}
	\psi(\lambda, n, \theta_1, t_1) &=\binom{n}{t_1} 
\frac{\left[\theta_1^{t_1} (1-\theta_1)^{n-t_1}\right]^{1+\lambda}}{{\left(\frac{c\log n}{n}\right)^{\lambda t_1}\left(1-\frac{c\log n}{n}\right)^{\lambda(n-t_1)} }}\text{,} \\
\zeta(k, \lambda, n, \boldtheta,  t_1)  &=
\exp\hspace{-0.5mm} \left(\lambda D_{1+\lambda}\Big(\hspace{-0.5mm}P_{Y^{n-t_1}|V=f(\boldtheta)} \big \| Q^{\ast(k-2)}_{Y^{n-t_1}}\hspace{-0.5mm} \Big)\hspace{-0.5mm}\right)\hspace{-0.5mm} \text{,}
\end{align}
and note that
\begin{align}
\zeta(k, \lambda, n, \boldtheta,  t_1)
&\le
C_1^\lambda(k-1)  \exp\left(\lambda \log (n-t_1)^{\frac{k-2}{2}} \right) \label{cant-be_saved} \\
&\le
C_1^\lambda(k-1) \exp\left(\lambda \log n^{\frac{k-2}{2}}\right)\text{,} \label{eqn:bunubunu}
\end{align}
where \eqref{cant-be_saved} follows from Lemma~\ref{lem:uniform_upper_bd_on_renyi_divergence_btw_model_and_jeff_mix} in Appendix~\ref{appdx:uniform upp bd on renyi div}. For $\boldtheta \in \cR_1$, 
\begin{align} 
&D_{1+\lambda }( P_{Y^n|V=\boldtheta} \| Q_{Y^n}^{\epsilon})  \nonumber \\
& \le \log \frac{k}{\epsilon} + D_{1+\lambda }\left( P_{Y^n|V=\boldtheta} \| M_1 \right) \label{eqn:general_ineq_here} \\
&=\log \frac{k}{\epsilon} + \frac{1}{\lambda}\log \sum_{t_1=0}^{n}  \psi (\lambda, n, \theta_1, t_1) \zeta(k, \lambda, n, \boldtheta,  t_1)  \\
&\le 
\log \frac{k}{\epsilon} + \log C_1(k-1) + \frac{k-2}{2} \log n \label{mertal2} \\ 
&\qquad \qquad \qquad \nonumber + D_{1+\lambda}\left(P_{Y^n|V=\theta_1} \| P_{Y^n|V=\frac{c\log n}{n}}\right)  
\end{align}
where \eqref{eqn:general_ineq_here} follows from \eqref{eqn:modified_jeffrey_k-1}, and \eqref{mertal2} is due to \eqref{eqn:bunubunu}. Finally, the desired result follows because \eqref{bbaa}--\eqref{bbac} imply
\begin{align}
	& D_{1+\lambda}\left( P_{Y^n|V=\theta_1}  \| P_{Y^n|V=\frac{c\log n}{n}}\right)\nonumber \\ 
	&\qquad = n D_{1+\lambda}\left( P_{Y|V=\theta_1}  \| P_{Y|V=\frac{c\log n}{n}}\right)  \\
	&\qquad \le \frac{c\log \rme}{1-\frac{c\log n}{n}} \log n \text{.}
\end{align}
\end{IEEEproof} 
It remains to investigate the supremization over $\cR_0$ in \eqref{eqn:max_of_sups_k-1}. Observe that
\begin{align}
&\sup_{\boldtheta \in \cR_0} D_{1+\lambda}(P_{Y^n|V=\boldtheta} \| Q_{Y^n}^\epsilon) \nonumber \\  
&\qquad  \le \log \frac{1}{1-\epsilon}+ \sup_{\boldtheta \in \cR_0} D_{1+\lambda}(P_{Y^n|V=\boldtheta} \| Q_{Y^n}^\ast )\text{,} \label{eqn:general-ilk-bd-before-prop2}
\end{align}
which follows from the definition of $Q_{Y^n}^\epsilon$ in \eqref{eqn:modified_jeffrey_k-1}. Parallel to Proposition~\ref{prop:achievability-k=2-hard-step}, Proposition~\ref{prop:general-hard to prove} characterizes the behavior of the supremum in the right side of \eqref{eqn:general-ilk-bd-before-prop2}.
\begin{prop} \label{prop:general-hard to prove}
For any $\lambda \in (0, \infty)$, 
\begin{align}
&\limsup_{n\to \infty} \left\{  \sup_{\boldtheta \in \cR_0} D_{1+\lambda}(P_{Y^n|V=\boldtheta}\| Q^{\ast}_{Y^n})-\frac{k-1}{2}\log \frac{n}{2\pi} \right\} \nonumber \\
&\qquad \le \log \frac{\opGamma^k(1/2)}{\opGamma(k/2)}-\frac{k-1}{2\lambda}\log(1+\lambda) \text{.} \label{aqzz}
\end{align}
\end{prop}
\begin{IEEEproof}
We are only interested in $\boldtheta \in \cR_0$. Therefore, for all $i=1, \hdots, k $,
\begin{align}
	\theta_i \ge \frac{c\log n}{n}\text{,}
\end{align}
where $c \in (0, 1/(2\log \rme))$ is a constant. Since there is an index $j$ such that $\theta_j\ge 1/k$, it simplifies notation without loss of generality that $j=k$. Otherwise, the proof remains identical. For a given positive integer $l$, let
\begin{align}
	\cI_l=\{i_1,\hdots, i_l \}\subset \cY
\end{align}
be a proper subset and note that
\begin{align}
&D_{1+\lambda}(P_{Y^n|V=\boldtheta}\| Q_{Y^n}^\ast) \nonumber \\
&\qquad =
\frac{1}{\lambda} \log( \mfrakV(k, \lambda, \boldtheta, n) + \mfrakW(k, \lambda, \boldtheta, n))\text{,} \label{aqzu}
\end{align}
where
\begin{align}
&\mfrakV(k, \lambda, \boldtheta, n)
= \sum_{l=1}^{k-1}\binom{k}{l} \sum_{\substack{\boldt \colon t_i \ge 0 \ \forall i \\ t_1+\cdots+ t_k=n \\  t_i=0 \ \forall i \in \cI_l}} \binom{n}{t_1 \cdots t_k}  \label{eqn:def:frakV-general} \\ 
&\quad \times (\theta_1^{t_1} \cdots \theta_k^{t_k})^{1+\lambda} \left( \frac{\opD_k(\frac12, \hdots, \frac12)}{\opD_k(t_1+\frac12, \hdots, t_k+\frac12)} \right)^\lambda  \text{,} \nonumber\\
&\mfrakW(k, \lambda, \boldtheta, n)=
\sum_{\substack{\boldt \colon t_i \ge 1 \ \forall i \\ t_1+\cdots+ t_k=n }} \binom{n}{t_1 \cdots t_k} \label{eqn:def:frakW-general}\\ 
&\quad \times (\theta_1^{t_1} \cdots \theta_k^{t_k})^{1+\lambda} \left(\frac{\opD_k(\frac12, \hdots, \frac12)}{\opD_k(t_1+\frac12, \hdots, t_k+\frac12)} \right)^\lambda\text{.}  \nonumber
\end{align}
Thanks to Lemma~\ref{lem:edge cases of ti} in Appendix~\ref{appdx:edge cases of t_i}, we know that for all sufficiently large $n$ satisfying
\begin{align}
	\frac{k\ln n}{2n}<1\text{,}
\end{align}
it follows that
\begin{align}
\mfrakV(k, \lambda, \boldtheta, n) \le \widetilde{C}(k,\lambda) n^{-(1+\lambda)c\log \rme } n^{\lambda \left(\frac{k-1}{2} 
\right)}\text{,} \label{eqn:thanks_to_thm5}
\end{align}
where $\widetilde{C}(k,\lambda)$ is a constant depending only on $\lambda$ and $k$, which is explicitly given in the proof of Lemma~\ref{lem:edge cases of ti}, see \eqref{eqn:C-tilde}. Hence, we may now focus attention on $\mfrakW(k, \lambda, \boldtheta, n) $. Note that
\begin{align}
\binom{n}{t_1 \cdots t_k} &\le \frac{\rme^{nH(\widehat{P}_{y^n})}}{(2\pi)^{\frac{k-1}{2}} } \left(\frac{n}{\prod_{i=1}^{k}t_i}\right)^{\frac12}
\rme^{\frac{1}{12n}}\text{,} \label{aqyx} 
\end{align}
and
\begin{align}
\prod_{i=1}^{k} \theta_i^{t_i}  &= \exp_\rme\hspace{-0.5mm} \left(-n \left[ D(\widehat{P}_{y^n}\| P_{Y|V=\boldtheta})+ H(\widehat{P}_{y^n}) \right] \right) \text{,} \label{aqyy} 
\end{align}
where both the entropy and relative entropy are in nats and the bound in \eqref{aqyx} follows from Stirling's approximation, see \eqref{eqn:upper bound on multinomial}. Note also that
\begin{align}
&\frac{1}{\opD_k(t_1+\frac12, \hdots, t_k+\frac12)} =
\frac{\opGamma(n+\frac k2) }{\prod_{i=1}^{k} \opGamma(t_i+\frac 12) }
\\
&\qquad \le \left( \frac{n}{2\pi} \right)^{\frac{k-1}{2}} \frac{\rme^{n H(\widehat{P}_{y^n})} \left( 1+\frac{k}{2n} \right)^{n+\frac{k-1}{2}}\rme^\frac{1}{12n+6k}}{\prod_{i=1}^{k}\left( 1+\frac{1}{2t_i} \right)^{ t_i }\left(2-\rme^{\frac{1}{12t_i+6}}\right)}\text{,} \label{aqyz}
\end{align}
where \eqref{aqyz} also follows from an application of Stirling's approximation, see \eqref{eqn:Stirling_Approximation_for_Gamma_Function}.

By substituting \eqref{aqyx}, \eqref{aqyy} and \eqref{aqyz} into the right side of \eqref{eqn:def:frakW-general}, we get
\begin{align}
\mfrakW(k, \lambda, \boldtheta, n) 
\le 
\left( \frac{n}{2\pi} \right)^{\frac{\lambda (k-1)}{2}} \opD_k^\lambda\Big(\tfrac12, \hdots, \tfrac12\Big) \mfrakS(k,\lambda, \boldtheta, n)\text{,} \label{eqn:thanks_to_substitition}
\end{align}
where
\begin{align}
&\mfrakS(k,\lambda, \boldtheta, n)=\hspace{-1mm} \sum_{\substack{\boldt \colon t_i \ge 1 \ \forall i \\ t_1+\cdots+ t_k=n }}  \hspace{-0.7mm} \frac{K(k,\lambda, n, \boldt)}{(2\pi)^{\frac{k-1}{2}} } \left(\frac{n}{\prod_{i=1}^{k}t_i}\right)^{\frac12} \label{eqn:def:frakS-general}\\ & \qquad
\times \exp_\rme\left(-n(1+\lambda)D(\widehat{P}_{y^n}\| P_{Y|V=\boldtheta})\right)\text{,} \nonumber
\end{align}
and
\begin{align}
K(k,\lambda, n, \boldt )&=\rme^{\frac{1}{12n}} \hspace{-1mm} \left(\frac{\left( 1+\frac{k}{2n} \right)^{n+\frac{(k-1)}{2}}\rme^\frac{1}{12n+6k}}{\prod_{i=1}^{k}\left( 1+\frac{1}{2t_i} \right)^{ t_i }\left(2-\rme^{\frac{1}{12t_i+6}}\right)}  \right)^\lambda \hspace{-1mm} \text{.} \label{eqn:def:K-general}
\end{align}
Observe once again that we can find an asymptotically suboptimal upper bound on $\mfrakS (k, \lambda, \boldtheta, n)$ that depends only on $k$ and $\lambda $ by invoking Lemma~\ref{lem:unif_upp_bd_on_K} in Appendix~\ref{appdx:bounds on K}, which shows a non-asymptotic uniform upper bound on $K(k, \lambda, n, \boldt)$, and then by invoking Lemma~\ref{lem:uniform bound on the sum} in Appendix~\ref{appdx:uniform upper bound on the sum}, which shows a non-asymptotic uniform upper bound on
\begin{align}
\mfrakT(k, \lambda, \boldtheta, n)&= \sum_{\substack{\boldt \colon t_i \ge 1 \ \forall i \\ t_1+\cdots+ t_k=n }}  \frac{1}{(2\pi)^{\frac{k-1}{2}}} \left(\frac{n}{\prod_{i=1}^{k}t_i}\right)^\frac12 \label{eqn:def:mfrakT-general}\\ 
&\qquad
\times \exp_\rme\left(-n(1+\lambda)D(\widehat{P}_{y^n}\| P_{Y|V=\boldtheta})\right)\text{.} \nonumber
\end{align}
Finding the optimal upper bound, on the other hand, requires a uniform Laplace approximation on $\mfrakS(k, \lambda, \boldtheta, n)$, which is introduced next. First, given $\delta\in (0,1/(k-1))$, recall the set $\cM_n$ as defined in \eqref{eqn:def:set M}, let 
\begin{align}
&\cN^\boldtheta_\delta = \cM_n \cap \left\{ (a_1, \hdots, a_k)\in \bbZ^k_+ \colon  \left|\frac{a_i}{n\theta_i}-1\right|\le \delta \ \forall i \right\}\text{,}\label{eqn:defn:N-delta}
\end{align}
and split $\mfrakS(k,\lambda, \boldtheta, n) $ as
\begin{align}
\mfrakS(k, \lambda, \boldtheta, n) = S_1(k, \lambda, \boldtheta,n,\delta)+S_2(k, \lambda, \boldtheta,n,\delta)\text{,} \label{eqn:S1_plus_S2}
\end{align} 
where
\begin{align}
S_1(k, \lambda, \boldtheta,n,\delta) &= \hspace{-0.4mm} \sum_{\substack{\boldt \colon \boldt \in \cN^\boldtheta_\delta\\ t_i\ge 1 \ \forall i}}  \frac{K(k,\lambda, n, \boldt)}{(2\pi)^{\frac{k-1}{2}} } \left(\frac{n}{\prod_{i=1}^{k}t_i}\right)^\frac12 \label{eqn:def:general-S1} \\
&\times \exp_\rme\left(-n(1+\lambda)D(\widehat{P}_{y^n}\| P_{Y|V=\boldtheta})\right)\text{,} \nonumber  \\
S_2(k, \lambda, \boldtheta,n,\delta) &= \hspace{-0.4mm} \sum_{\substack{\boldt \colon \boldt \not\in \cN^\boldtheta_\delta \\ t_i\ge 1 \ \forall i}}  \frac{K(k,\lambda, n, \boldt)}{(2\pi)^{\frac{k-1}{2}} } \left(\frac{n}{\prod_{i=1}^{k}t_i}\right)^\frac12 \label{eqn:def:general-S2} \\  
&\times \exp_\rme\left(-n(1+\lambda)D(\widehat{P}_{y^n}\| P_{Y|V=\boldtheta})\right)\text{.}  \nonumber
\end{align}
In Lemmas~\ref{lem:generela_sum_S1} and~\ref{lem:generela_sum_S2} in Appendix~\ref{appdx:achievability_k>2}, we show that the following properties hold:
\begin{align}
\lim_{\delta \to 0}\limsup_{n\to \infty} \sup_{\substack{\boldtheta \in \cR_0 \\ \theta_k \ge 1/k}} S_1(k, \lambda, \boldtheta,n,\delta)  &\le (1+\lambda)^{-\frac{k-1}{2}}\text{,} \label{eqn:reslt:general-S1} \\
\lim_{n \to \infty}\sup_{\substack{\boldtheta \in \cR_0 \\ \theta_k \ge 1/k}} S_2(k, \lambda, \boldtheta,n,\delta) &=0 \quad \forall \delta \in (0,1) \text{.} \label{eqn:reslt:general-S2}
\end{align}	
Since the left side of \eqref{eqn:S1_plus_S2} does not depend on $\delta $, \eqref{eqn:reslt:general-S1} and \eqref{eqn:reslt:general-S2} imply, by letting $\delta \to 0 $, that 
\begin{align}
\limsup_{n\to \infty } \sup_{\substack{ \boldtheta \in \cR_0 \\ \theta_k \ge 1/k }} \mfrakS(k,\lambda, \boldtheta, n) \le (1+\lambda)^{-\frac{k-1}{2}}\text{.} \label{eqn:last-cmt-limit-general}
\end{align}
Finally, it follows from (\ref{aqzu}), (\ref{eqn:thanks_to_thm5}), (\ref{eqn:thanks_to_substitition}), and \eqref{eqn:last-cmt-limit-general}, that \eqref{aqzz} holds when $\theta_k \ge 1/k $ as we wanted to show.
\end{IEEEproof} 
Invoking Proposition~\ref{prop:achievability-general-faces dont matter}, we see that for each $i=1, \hdots, k $
\begin{align}
&\sup _{\boldtheta \in \cR_i}  D_{1+\lambda }\left( P_{Y^n|V=\boldtheta} \| Q_{Y^n}^{\epsilon}\right) - \frac{k-1}{2}\log \frac{n}{2\pi} \nonumber \\
&\le \bigg(\frac{c\log \rme }{1-\frac{c\log n }{n}}-\frac{1}{2} \bigg)\log n +\frac{k-1}{2}\log (2\pi)  +\log \frac{k}{\epsilon} \label{eqn:general_being_dominated}\\ 
&\qquad + \log C_1(k-1)  \text{,} \nonumber
\end{align}
while thanks to \eqref{eqn:general-ilk-bd-before-prop2} and Proposition~\ref{prop:general-hard to prove}, it follows that
\begin{align}
&\limsup_{n \to \infty } \left\{ \sup_{\boldtheta \in \cR_0 } D_{1+\lambda}(P_{Y^n | V=\boldtheta }\| Q_{Y^n}^\epsilon )-\frac{k-1}{2}\log \frac{n}{2\pi} \right\} \nonumber \\ 
&\qquad \le  \log \frac{\opGamma^k(1/2) }{\opGamma(k/2) }-\frac{k-1}{2\lambda} \log(1+\lambda) +\log \frac{1}{1-\epsilon }\text{.} \label{eqn:general-dominating}
\end{align} 
Since $c \in (0, 1/(2\log \rme)) $, we see that, as $n \to \infty $, the right side of \eqref{eqn:general_being_dominated} goes to $-\infty $ whereas the right side of \eqref{eqn:general-dominating} remains constant. In view of \eqref{eqn:max_of_sups_k-1}, \eqref{eqn:general_being_dominated} and \eqref{eqn:general-dominating}, the desired result in \eqref{eqn:achievability-to-show-general} follows by choosing an arbitrarily small $\epsilon $ in \eqref{eqn:mod_jeff_prior_k-1}.  \hfill \IEEEQED

\begin{appendices}
%%%%%%%%%%
\section{Explicit Evaluation of $\alpha$-Mutual Information}\label{appdx: Explicit alpha mutual info}

In the case of finite collection of arbitrary distributions, explicit evaluation of $I_{1+\lambda}$ is provided by Sibson \cite[Corollary 2.3]{sibson1969information}. A more general result that allows non-discrete alphabets can be found in \cite{Verdu2015}.
\begin{lem}\label{lem:explicit alpha mut info}
	Let $\lambda \in (0,\infty) $. Given an arbitrary input distribution $P_V$ on $\bold{\Theta}$ and a random transformation $P_{Y|V}\colon \bold{\Theta} \to \cY  $ with finite output alphabet $\cY$, the $\alpha$-mutual information of order $1+\lambda$ induced by $P_V$ on $P_{Y|V} $ satisfies
		\begin{align}
		&\frac{\lambda}{1+\lambda} I_{1+\lambda}(P_V,P_{Y|V}) \nonumber \\
		 &\qquad = \log \sum_{y\in \cY} \left(  \int_{\boldtheta \in \bold{\Theta}} P^{1+\lambda}_{Y|V=\boldtheta}(y)\rmd P_V(\boldtheta)\right)^{\frac1{1+\lambda}}\text{.} \label{sibson_lem_res}
	\end{align}
\end{lem}

\begin{IEEEproof}
Define 
\begin{align}
	R_Y(y)=\frac{\left( \ds \int_{\boldtheta\in \bold{\Theta}} P^{1+\lambda}_{Y|V=\boldtheta}(y)\rmd P_V(\boldtheta)  \right)^{\frac{1}{1+\lambda}} }{\ds \sum_{b\in \cY} \left( \int_{\boldxi \in \bold{\Theta }}P^{1+\lambda}_{Y|V=\boldxi}(b)\rmd P_V(\boldxi)  \right)^{\frac{1}{1+\lambda}} } \text{,}
\end{align}
and recall that 
\begin{align}
	D_{1+\lambda}(R_Y\|Q_Y)\ge 0 \label{yucely}
\end{align}
for any distribution $Q_Y$ on $\cY$. Capitalizing on \eqref{yucely}, note that 
\begin{align}
&D_{1+\lambda}(P_{Y|V}P_V \| Q_Y P_V) \nonumber \\
	&\quad =\frac{1}{\lambda} \log \sum_{y\in \cY}  \int_{\boldtheta \in\bold{\Theta} } \frac{P^{1+\lambda}_{Y|V=\boldtheta}(y)}{Q_Y^{\lambda}(y)} \rmd P_V(\boldtheta) \label{yucelz}
	\\ 
	&\quad \ge \frac{1+\lambda}{\lambda} \log \sum_{b\in \cY}\left(\int_{\boldxi \in \bold{\Theta} } P^{1+\lambda}_{Y|V=\boldxi}(b)\rmd P_{V}(\boldxi)\right)^{\frac{1}{1+\lambda}}  \\
	&\quad = D_{1+\lambda}(P_{Y|V}P_V \| R_Y P_V ) \label{yucel} \text{.}
\end{align}
By the definition of the $\alpha$-mutual information, see \eqref{eqn:alpha-mutual-information definition}; \eqref{yucel} implies the result in \eqref{sibson_lem_res}.
\end{IEEEproof}

\section{Monotonicity of Binary R\'enyi Divergence} \label{appdx:monotonicity of binary renyi divergence}

\begin{lem} \label{lem:renyi divergence monotone decreasing}
Let $P_{Y|V=\theta}$ denote a Bernoulli distribution with parameter $\theta$. For any $\xi \in (0,1]$ and $\lambda \in (0,\infty) $, $D_{1+\lambda}(P_{Y|V=\theta}\|P_{Y|V=\xi})$ is a monotone decreasing function of $\theta$ on $[0,\xi]$. 
\end{lem}
\begin{IEEEproof} 
Fix $\lambda \in (0, \infty)$. Let $Y\sim P_{Y|V=\theta}$. It suffices to prove that $\bbE\left[ \left( \frac{P_{Y|V=\theta}(Y)}{P_{Y|V=\xi}(Y)} \right)^\lambda \right]  $
is a monotone decreasing function of $\theta$ on $[0,\xi]$. To that end, note that
\begin{align}
\frac{\rmd}{\rmd\theta}\bbE\bigg[ \bigg( \frac{P_{Y|V=\theta}(Y)}{P_{Y|V=\xi}(Y)} \bigg)^\lambda \bigg] \hspace{-0.3mm}
&=\hspace{-0.3mm} (1+\lambda) \bigg( \frac{\theta^{\lambda}}{\xi^{\lambda}}-\frac{(1-\theta)^{\lambda}}{(1-\xi)^{\lambda}} \bigg) \\
&\le 0\text{,} \label{eqn:see_key}
\end{align}
where \eqref{eqn:see_key} follows because $\theta\in [0,\xi]$ implies
\begin{align}
\frac{\theta}{\xi} \le \frac{1-\theta}{1-\xi}\text{.}
\end{align} 
\end{IEEEproof}
%%%%%%%%%
\section{Uniform Upper Bound on $D_{1+\lambda }\left( P_{Y^n|V=\boldtheta} \| Q_{Y^n}^{\ast}\right)$} \label{appdx:uniform upp bd on renyi div}
\begin{lem} \label{lem:uniform_upper_bd_on_renyi_divergence_btw_model_and_jeff_mix}
Let $\boldtheta \in \simplex^{k-1}$ be an element in the $(k-1)$-dimensional simplex and assume that we are given a discrete i.i.d. model $P_{Y^n|V=\boldtheta}$. Then, for any $n\ge 1$ and $y^n \in \cY^n$, the relative information between the model $P_{Y^n|V=\boldtheta}$ and Jeffreys' mixture $Q_{Y^n}^{\ast}$ satisfies the following bound
\begin{align}
\imath_{P_{Y^n|V=\boldtheta}\| Q_{Y^n}^{\ast}} (y^n)
\le
 \frac{k-1}{2}\log n +\log C_1(k)\text{,}
\end{align}
where 
\begin{align}
C_1(k)=
\frac{\rme^{\frac{6k+1}{12}}\opD_k\left(\frac{1}{2}, \cdots, \frac{1}{2}\right)}{
\left(2\pi\right)^{\frac{k-1}{2}}\left(2-\rme^{1/6}\right)^k  } \left(1+\frac{k}{2} \right)^{\frac{k-1}{2}}\text{.} \label{eqn:C_1(k)}
\end{align}
Consequently, for any $\lambda > 0 $,
\begin{align}
D_{1+\lambda }\left( P_{Y^n|V=\boldtheta} \| Q_{Y^n}^{\ast}\right)
\le 
\frac{k-1}{2}\log n + \log C_1(k)\text{,}
\end{align}
where $C_1(k)$ is given in \eqref{eqn:C_1(k)}.
\end{lem}

\begin{IEEEproof}
Immediate consequence of \cite[Lemma 4]{XieBarron1997}.
\end{IEEEproof}
%%%%%%%%%%%
\section{Edge Cases of $t_i$}\label{appdx:edge cases of t_i}
\begin{lem}\label{lem:edge cases of ti}
Let $c\in (0, 1/(2\log \rme))$ and for a given positive integer $l$, let $\cI_l=\{i_1, \hdots, i_l\}$ be a proper subset of $\cY$. Then, for any $n$ satisfying
\begin{align}
\frac{k\ln n}{2n} < 1\text{,}
\end{align}
and $\boldtheta \in \cR_0$ (defined in \eqref{eqn:definition_of_R})
\begin{align}
\mfrakV(k, \lambda, \boldtheta, n) \le 
\widetilde{C}(k,\lambda) n^{-(1+\lambda)c\log \rme + \lambda \left(\frac{k-1}{2} 
\right)} \text{,} \label{eqn:edge cases of t_i bound}
\end{align} 
where $\mfrakV(k, \lambda, \boldtheta, n)$ is defined\footnote{The quantity $\mfrakV( \lambda, \theta_1, n) $ defined in \eqref{eqn:def:frakV-k=2} corresponds to the special case of \eqref{eqn:def:frakV-general} where $k=2 $, $\boldtheta=(\theta_1, 1-\theta_1 ) $.} in \eqref{eqn:def:frakV-general} and $\widetilde{C}(k,\lambda)$ is a constant that only depends $k$ and $\lambda$.
\end{lem}
\begin{IEEEproof}
Denote 
\begin{align}
\{i_{l+1}, i_{l+2}, \hdots, i_k\} = \{1, \hdots, k\} -\{i_1, i_2, \hdots, i_l\}\text{,}
\end{align}
and note that
\begin{align}
&\hspace{-3mm} \sum_{\substack{\boldt \colon t_i=0\ \forall i \in \cI_l \\ t_1+\cdots+t_k=n }} \hspace{-1.5mm}  \binom{n}{t_1 \cdots t_k} \hspace{-0.1mm} (\theta_1^{t_1} \cdots \theta_k^{t_k})^{1+\lambda} \frac{\opD^{\lambda}_k( \frac12, \ldots, \frac12)}{{\opD^{\lambda}_k(t_1 \hspace{-0.5mm} +\hspace{-0.5mm}\frac12, \ldots ,t_k\hspace{-0.5mm}+\hspace{-0.5mm}\frac12)}} \nonumber \\
&=
\sum_{\substack{t_{i_{l+1}},\hdots, t_{i_k} \\ t_{i_{l+1}}+\cdots+t_{i_k}=n }}  \binom{n}{t_{i_{l+1}}\cdots t_{i_k}}\left( \theta_{i_{l+1}}^{t_{i_{l+1}}} \cdots \theta_{i_k}^{t_{i_k}} \right)^{1+\lambda} \label{cczz:summation} \\
&\qquad \times \frac{\opD^{\lambda}_k (\frac12, \ldots, \frac12)}{{\opD^{\lambda}_k (\frac12, \hdots, \frac12, t_{i_{l+1}}+\frac12, \ldots ,t_{i_k}+\frac12 ) } }  \text{.} \nonumber
\end{align}
Regarding the last term within the summation in the right side of \eqref{cczz:summation},
\begin{align}
&\frac{\opD_k(\frac12, \ldots, \frac12) }{{\opD_k(\frac12, \hdots, \frac12, t_{i_{l+1}}+\frac12, \ldots ,t_{i_k}+\frac12 ) } } \nonumber \\
&=
\frac{\opD_{k-l}(\frac12, \hdots, \frac12)}{\opD_{k-l}(t_{i_{l+1}}+\frac12, \hdots, t_{i_k}+\frac12)}\frac{ \opGamma(\frac{k-l}{2})}{ \opGamma(\frac{k}{2})} 
\frac{ \opGamma(n+\frac{k}{2})}{ \opGamma(n+\frac{k-l}{2})} \label{eqn:prev_inequl}\\
&\le
\frac{\opD_{k-l}(\frac12, \hdots, \frac12)}{\opD_{k-l}(t_{i_{l+1}}+\frac12, \hdots, t_{i_k}+\frac12)}\frac{\opGamma(\frac{k-1}{2})}{\opGamma(\frac{k}{2})} 
\frac{\opGamma(n+\frac{k}{2})}{\opGamma(n+\frac{k-l}{2})}\text{,} \label{ccaa} 
\end{align}
where \eqref{eqn:prev_inequl} follows from the definition of the Dirichlet integrals in \eqref{eqn:def:dirichlet integrals}, and \eqref{ccaa} follows from the fact that $l\ge 1$. Now, observe that
\begin{align}
\frac{ \opGamma\left(n+\frac{k}{2}\right)}{\opGamma\left(n+\frac{k-l}{2}\right)} 
&=
\frac{\left(n+\frac{k}{2}\right)^{n+\frac{k-1}{2}}\rme^{-n-\frac{k}{2}}(1+r_0)}{\left(n+\frac{k-l}{2}\right)^{n+\frac{k-l-1}{2}}\rme^{-n-\frac{k-l}{2}}(1+r_l)} \label{ccab:stirling} \\
&\le 
\left(\frac{ \left(1+\frac{k}{2}\right)^{\frac{k-1}{2}} \rme^{k/2}\rme^{1/24}}
{ 2-\rme^{1/18}}
\right) n^{l/2}\text{,} \label{ccac}
\end{align} 
where $r_l$ is the remainder in Stirling's approximation of $\opGamma \left(n+\frac{k-l}{2} \right) $ in \eqref{eqn:Stirling_Approximation_for_Gamma_Function}, and \eqref{ccac} is due to the following elementary bounds:
\begin{align}
\left(1+\frac{k}{2n}\right)^{n+\frac{k-1}{2}} &\le  \left(1+\frac{k}{2}\right)^{\frac{k-1}{2}}\rme^{k/2}\text{,} \\
1+r_0 & \le \rme^{1/24}\text{,} \\
\left(1+\frac{k-l}{2n}\right)^{n+\frac{k-l-1}{2}} \rme^{l/2}&\ge  1\text{,} \\
1+r_l 
&\ge  2-\rme^{1/18}\text{.}
\end{align}
It follows that
\begin{align}
 & \frac{\opD_k(\frac12, \ldots, \frac12) }{{\opD_k(\frac12, \hdots, \frac12, t_{i_{l+1}}+\frac12, \ldots ,t_{i_k}+\frac12)}} \nonumber \\ 
&\qquad \le \frac{\opD_{k-l}(\frac12, \hdots, \frac12)}{\opD_{k-l}(t_{i_{l+1}}+\frac12, \hdots, t_{i_k}+\frac12)} 
C_2(k) n^{ l/2}\text{,} \label{ccad}
\end{align}
where
\begin{align}
C_2(k)= \frac{\opGamma(\frac{k-1}{2})}{ \opGamma(\frac{k}{2})}  
\frac{ (1+\frac{k}{2})^{\frac{k-1}{2}} \rme^{\frac{12k+1}{24}}}
{ 2-\rme^{1/18}} \text{.} \label{eqn:C_2}
\end{align}
Since $\boldtheta \in \cR_0$,
\begin{align}
(1-(\theta_{i_1}+\cdots + \theta_{i_l}))^n &\le 	\left(1-\frac{lc\log n}{n}\right)^{n}  \label{snap1} \\
&\le 
n^{-lc\log \rme}\text{,} \label{snap2} 
\end{align} 
where \eqref{snap2} is because $\frac{lc\log n}{n}	< \frac{k \ln n }{2n} < 1$ and for any $x<1$ we have $\log (1-x)\le -x\log \rme$. Let
\begin{align}
\bar{\boldtheta}&=\left(\bar{\theta}_{i_{l+1}}, \hdots, \bar{\theta}_{i_k} \right) \\
&=\frac{\left( \theta_{i_{l+1}}, \cdots,  \theta_{i_k} \right)}{1-(\theta_{i_1}+\cdots + \theta_{i_l})} \text{.} \label{eqn:bar_theta}
\end{align}
It follows from \eqref{ccad} and \eqref{snap2} that
\begin{align}
&\sum_{\substack{t_{i_{l+1}},\hdots, t_{i_k} \\ t_{i_{l+1}}+\cdots+t_{i_k}=n }}  \binom{n}{t_{i_{l+1}}\cdots t_{i_k}}\left( \theta_{i_{l+1}}^{t_{i_{l+1}}} \cdots \theta_{i_k}^{t_{i_k}} \right)^{1+\lambda} \nonumber \\
&\qquad \qquad \quad \times \frac{\opD^{\lambda}_k\left(\frac12, \ldots, \frac12\right) }{{\opD^{\lambda}_k\left(\frac12, \hdots, \frac12, t_{i_{l+1}}+\frac12, \ldots ,t_{i_k}+\frac12\right)}} \nonumber  \\
&\le
\sum_{\substack{t_{i_{l+1}},\hdots, t_{i_k} \\ t_{i_{l+1}}+\cdots+t_{i_k}=n }}  \binom{n}{t_{i_{l+1}}\cdots t_{i_k}}\left(\bar{\theta}_{i_{l+1}}^{t_{i_{l+1}}} \cdots \bar{\theta}_{i_k}^{t_{i_k}} \right)^{1+\lambda} \label{cdaa}\\
& \qquad \qquad \quad \times \frac{\opD^\lambda_{k-l}(\frac12, \hdots, \frac12)}{\opD^\lambda_{k-l}(t_{i_{l+1}}+\frac12, \hdots, t_{i_k}+\frac12)}
 \frac{C_2^\lambda(k) n^{ \lambda l/2}}{n^{(1+\lambda)lc \log \rme}}.   \nonumber
\end{align} 
Note that
\begin{align}
&\exp\left(\lambda D_{1+\lambda}\left( P_{Y^n|V=\bar{\boldtheta}} \big\| Q_{Y^n}^{\ast{(k-l-1)}}\right)\right) \nonumber
  \\
&= \sum_{\substack{t_{i_{l+1}},\hdots, t_{i_k} \\ t_{i_{l+1}}+\cdots+t_{i_k}=n }}  \binom{n}{t_{i_{l+1}}\cdots t_{i_k}}\left(\bar{\theta}_{i_{l+1}}^{t_{i_{l+1}}} \cdots \bar{\theta}_{i_k}^{t_{i_k}} \right)^{1+\lambda}
\\  
&\qquad \qquad \quad  \times 
\frac{\opD^\lambda_{k-l}(\frac12, \hdots, \frac12)}{\opD^\lambda_{k-l}(t_{i_{l+1}}+\frac12, \hdots, t_{i_k}+\frac12)}
\text{,} \nonumber
\end{align}	
where $Q_{Y^n}^{\ast{(k-l-1)}}$ denotes the Jeffreys' mixture when the underlying parameter space is the $(k-l-1)$-dimensional simplex. Using the uniform upper bound on R\'enyi divergence in Lemma~\ref{lem:uniform_upper_bd_on_renyi_divergence_btw_model_and_jeff_mix}, we get
\begin{align}
\exp \hspace{-0.5mm}\big(\lambda D_{1+\lambda}\big( P_{Y^n|V=\bar{\boldtheta}} \| Q_{Y^n}^{\ast{(k-l-1)}}\big)\big)  \le C_1^\lambda(k-l) n^{\lambda(\frac{k-l-1}{2})}\text{,}
\end{align}	
where $C_1(k)$ is as defined in \eqref{eqn:C_1(k)}. Since $l\in \{1, \hdots, k-1 \} $, $\opD_1(\frac12)=1 $, and $\opD_m(\frac12, \hdots, \frac12)\le \pi $ for any integer $m\ge 2 $, we can upper bound
\begin{align}
C_1(k-l)&\le 
\frac{\ds  \pi\rme^{\frac{6k-5}{12}}}{
\ds 2-\rme^{1/6}  } \left(1+\frac{k-1}{2} \right)^{\frac{k-2}{2}} \label{eqn:bound_on_C_1(k-l)} \\
&=C_3(k)\text{.} \label{eqn:C_3}
\end{align}	
As a result,
\begin{align}
&\mfrakV(k, \lambda, \boldtheta, n) \nonumber \\
&\qquad \le 
\sum_{l=1}^{k-1}\binom{k}{l}C_2^\lambda(k) C_3^\lambda(k) n^{-(\lambda+1)c \log \rme+\lambda \left(\frac{k-1}{2} \right)}  \\
&\qquad =
(2^k-2)C_2^\lambda(k) C_3^\lambda(k) n^{-(\lambda+1)c \log \rme+\lambda \left(\frac{k-1}{2} \right)} \text{,}
\end{align}
and \eqref{eqn:edge cases of t_i bound} follows after setting
\begin{align}
\widetilde{C}(k,\lambda)=(2^k-2)C_2^\lambda(k) C_3^\lambda(k)\text{.} \label{eqn:C-tilde}
\end{align} 	
\end{IEEEproof}
%%%%%%%%
\section{Uniform Upper Bound on $\mfrakT(k, \lambda, \boldtheta, n) $ } \label{appdx:uniform upper bound on the sum}
The quantity defined\footnote{The quantity $\mfrakT(\lambda, \theta_1, n) $ defined in \eqref{eqn:def:mfrakT-k=2} corresponds to the special case of \eqref{eqn:def:mfrakT-general} where $k=2 $, $\boldtheta=(\theta_1, 1-\theta_1 ) $.} in \eqref{eqn:def:mfrakT-general} satisfies the following upper bound.
\begin{lem} \label{lem:uniform bound on the sum}
\begin{align}
\mfrakT(k, \lambda, \boldtheta, n) \le \frac{C_1^\lambda(k) (2\pi)^{\frac{\lambda(k-1)}{2}} }{\opD_k^\lambda(1/2, \hdots, 1/2)}\frac{\rme^\frac{k(20\lambda+3)}{36}}{(2-\rme^\frac{1}{6k})^\lambda}\text{,} \label{eqn:lem:mfrakT}
\end{align}
where $C_1(k)$ is explicitly given in \eqref{eqn:C_1(k)}.
\end{lem}
\begin{IEEEproof}
Define
\begin{align}
\widetilde{\mfrakS}(k, \lambda, \boldtheta, n ) &= \hspace{-0.8mm} \sum_{\substack{\boldt \colon t_i\ge 1 \ \forall i \\ t_1+\cdots+t_k=n} } \hspace{-0.8mm} \frac{\widetilde{K}(k, \lambda, n, \boldt )}{(2\pi)^\frac{k-1}{2}} \left(\frac{n}{\prod_{i=1}^{k}t_i}\right)^\frac12  \\
&\qquad \times \exp_\rme(-n(1+\lambda)D(\widehat{P}_{y^n}\|P_{Y|V=\boldtheta}))\text{,} \nonumber 
\end{align}
where
\begin{align}
\widetilde{K}(k, \lambda, n, \boldt ) &=
\frac{\rme^\frac{1}{12(n+1)}}{\rme^\frac{k}{12}}\left( \frac{\left( 1+\frac{k}{2n} \right)^{n+\frac{k-1}{2}}}{ \prod_{i=1}^{k}\left( 1+\frac{1}{2t_i} \right)^{t_i}} \right)^\lambda  \\ 
&\qquad \times
\left(\frac{2-\rme^{\frac{1}{12n+6k}}}{\prod_{i=1}^{k} \rme^\frac{1}{12t_i+6}}\right)^\lambda\text{.} \nonumber
\end{align}
Note that
\begin{align}
&D_{1+\lambda}(P_{Y^n|V=\boldtheta}\|Q^\ast_{Y^n}) \ge
\frac{1}{\lambda}\log \mfrakW(k, \lambda, \boldtheta, n) \label{after_psi-shorter} \\
&\ge 
\frac{1}{\lambda}\log  \Big(\hspace{-0.1mm}\Big(\frac{n}{2\pi} \Big)^\frac{\lambda(k-1)}{2} \opD^\lambda_k(\tfrac12, \hdots, \tfrac12)\widetilde{\mfrakS}(k, \lambda, \boldtheta, n )\hspace{-0.1mm} \Big)  \text{,} \label{bxzy}
\end{align}
where \eqref{after_psi-shorter} follows from \eqref{aqzu}, and \eqref{bxzy} follows from Stirling's approximations, \eqref{eqn:lower bound on multinomial} and \eqref{eqn:Stirling_Approximation_for_Gamma_Function}, as well as the fact that
\begin{align}
\prod_{i=1}^{k} \theta_i^{t_i} =\exp_\rme \left(-n\left[D(\widehat{P}_{y^n}\|P_{Y|V=\boldtheta})+H(\widehat{P}_{y^n})\right]\right) \text{.}
\end{align}
Regarding $\widetilde{K}(k, \lambda, n, \boldt )$, one can check that
\begin{align}
\widetilde{K}(k, \lambda, n, \boldt )\ge \frac{(2-\rme^\frac{1}{6k})^\lambda}{\rme^{\frac{k(20\lambda+3)}{36}}} \text{.} \label{bound-for-tilde-K}
\end{align}
Invoking Lemma~\ref{lem:uniform_upper_bd_on_renyi_divergence_btw_model_and_jeff_mix} in Appendix~\ref{appdx:uniform upp bd on renyi div} to upper bound the left side of \eqref{after_psi-shorter} and applying the bound in \eqref{bound-for-tilde-K} to \eqref{bxzy} results in \eqref{eqn:lem:mfrakT}.
\end{IEEEproof}
%%%%%%%%%%%%%
\section{Bounds on $K(k,\lambda, n, \boldt )$} \label{appdx:bounds on K}
The quantity defined\footnote{The quantity $K(\lambda, n, t_1)$ defined in \eqref{eqn:def:K-when_k=2} corresponds to the special case of \eqref{eqn:def:K-general} where $k=2$, $\boldt=(t_1, n-t_1)$.} in \eqref{eqn:def:K-general} satisfies the following non-asymptotic bound.
\begin{lem}[Uniform Upper Bound on $K(k,\lambda, n, \boldt )$]	\label{lem:unif_upp_bd_on_K}
Given $\lambda\in (0,  \infty)$,
\begin{align}
K(k,\lambda, n, \boldt )&\le \rme^\frac{1}{12}\left( \frac{ \rme^{\frac{k}{2}}\big( 1+\frac{k}{2} \big)^{\frac{k-1}{2}} \rme^\frac{1}{12+6k}      }{\big( \frac{3}{2} ( 2-\rme^\frac{1}{18} )\big)^k } \right)^\lambda \label{aqwx} \\
&=M(k, \lambda)\text{.}
\end{align}
In particular,
\begin{align}
K(\lambda, n, t_1) &\le 	M(2, \lambda) \\
&\le 3^\lambda\rme^\frac{1}{12} \label{eqn:M(2, lambda)le}
\end{align}
\end{lem}
\begin{IEEEproof}
For $x\ge 1$,
\begin{align}
\left( 1+\frac{1}{2x} \right)^{x}  \left( 2-\rme^{\frac{1}{12x+6}} \right) \ge \frac{3}{2}\left(2- \rme^{\frac{1}{18}} \right)\text{,} \label{appi:func}
\end{align}
because the function in the left side of \eqref{appi:func} is an increasing function. On the other hand,
\begin{align}
\left( 1+\frac{k}{2x} \right)^{\frac{k-1}{2}}\rme^{\frac{1}{12x+6k}}\le \left(1+\frac{k}{2} \right)^\frac{k-1}{2}\rme^\frac{1}{12+6k}\text{,} \label{appi_func2}
\end{align}
because the function of the left side of \eqref{appi_func2} is a decreasing function. Finally, \eqref{aqwx} follows from the fact that $\lambda \ge 0 $ and $\rme^\frac{k}{2}  \ge \left( 1+\frac{k}{2n} \right)^{n}$. 
\end{IEEEproof}
\begin{lem}[Asymptotic Upper Bound on $K(k, \lambda, n, \boldt )$] \label{lem:limit_of_K}
Let $c \in (0, 1/(2\log \rme ) ) $, and $\delta\in(0,1/(k-1)) $ be fixed and $n>2$ be an integer. Assume that $\boldtheta\in \cR_0 $ (defined in \eqref{eqn:definition_of_R}) satisfies $\theta_k \ge 1/k$. If for $i\in \{1,\hdots, k-1 \}$
\begin{align}
 n(1-\delta)\theta_i  \le t_i \le  n(1+\delta)\theta_i  \text{,}
\end{align}
then
\begin{align}
K(k, \lambda, n, \boldt ) &\le  K(k,\lambda, n, c(1-\delta)\mathbold{u} ) \label{ceaa} \\
&=M(k, \lambda, n, c, \delta)\text{,}
\end{align}
where in \eqref{ceaa} $\mathbold{u} = (u_1, \ldots, u_k) \in \bbR^k $ satisfies
\begin{align}
\mathbold{u} = \left(\log n  , \hdots, \log n, \frac{(1-(k-1)\delta)n}{c(1-\delta)k}\right)	\text{.}
\end{align}
Furthermore,
\begin{align}
\lim_{n\to \infty} M(k, \lambda, n, c, \delta) = 1\text{.} 
\label{ceab}
\end{align}
\end{lem}
\begin{IEEEproof}
Note that since $\boldtheta\in \cR_0$, for $i\in \{1,\hdots, k \}$
\begin{align}
t_i &\ge c(1-\delta) u_i \\
&= v_i \text{,} 
\end{align}
which, in turn, imply that 
\begin{align}
\left( 1+\frac{1}{2t_i} \right)^{t_i} &\ge \left(1+\frac{1}{2{v_i}} \right)^{{v_i}}  \label{apph:func1} \\
\rme^{\frac{1}{12t_i+6}}   &\le  \rme^{\frac{1}{12{v_i}+6}}  \text{.}
\end{align}
Hence, inequality \eqref{ceaa} follows. It is straightforward to see the limit in \eqref{ceab}.
\end{IEEEproof}
%%%%%%%%%%%%
\section{Lemmas for the Proof in Section~\ref{sec:subsec:ach-k=2}} \label{appdx:achievebility_k=2}
In the proofs of Lemmas~\ref{lem:k=2_sum_S1_asympt},~\ref{lem:k=2_sum_S2_asympt} and~\ref{lem:k=2_sum_S3_asympt}, we use the following bound: for $\theta \in (0, 1/2)$ and $\delta\in (0,1)$,
\begin{align}
|\tau-\theta|\le \delta \theta \implies d(\tau\| \theta)\ge \frac{1}{2} \frac{(1-\delta)(\tau-\theta)^2}{\theta(1-\theta)}\text{,} \label{taylor-ineq-k=2}
\end{align}
in nats. In particular, when $0< \tau \le \theta \le 1/2 $
\begin{align}
	d(\tau\|\theta)\ge \frac{1}{2} \frac{(\tau-\theta)^2}{\theta(1-\theta)}\text{.} \label{taylor-ineq-k=2_no2}
\end{align}
To show \eqref{taylor-ineq-k=2} and \eqref{taylor-ineq-k=2_no2}, we rely on Taylor's theorem:
\begin{align}
d(\tau\| \theta)=\frac{1}{2}\frac{(\tau-\theta)^2}{\theta(1-\theta)}+\frac{2\alpha-1}{6\alpha^2(1-\alpha)^2}(\tau-\theta)^3\text{,} \label{taylor-eq-k=2}
\end{align}
for some $\alpha$ in between $\tau$ and $\theta$. 

\begin{lem}\label{lem:k=2_sum_S1_asympt}
Let $c\in (0, 1/(2\log \rme))$ and fix $\delta \in (0,1)$.
\begin{align}
\lim_{n \to \infty} \sup_{\theta_1\in \left[ \frac{c\log n}{n}, \frac{1}{2} \right]} S_1(\lambda, \theta_1, n, \delta)=0\text{,} \label{eqn:lem:achiv-k=2-S1}
\end{align}
where $S_1(\lambda, \theta_1, n, \delta)$ is defined in \eqref{eqn:k=2_sum_S1}.
\end{lem}
\begin{IEEEproof}
Assume that $n$ is a sufficiently large integer, let $\theta_1\in \left[ \frac{c\log n}{n}, \frac{1}{2} \right] $ be given. Then
\begin{align}
& S_1(\lambda, \theta_1, n, \delta) \nonumber \\
&\le 
\sum_{t_1=1}^{\lfloor n(1-\delta)\theta_1 \rfloor} \left(\frac{9^\lambda \rme^\frac16 n}{2\pi t_1t_2}\right)^\frac12 \exp_\rme\left(-\frac{1}{2}n(1+\lambda) {\delta^2 \theta_1}\right)\label{appi:app_lem}   \\
&\le 
 (1-\delta)\theta_1 \hspace{-0.5mm} \left(\hspace{-0.5mm}\frac{9^\lambda \rme^\frac16 n^3}{2\pi(n-1)}\hspace{-0.5mm}\right)^{\frac12} \exp_\rme\hspace{-0.5mm} \left(-\frac{n}{2}(1+\lambda)  {\delta^2 \theta_1}\right)\hspace{-0.5mm} \text{,} \label{appi:final}
\end{align}
where \eqref{appi:app_lem} is due to \eqref{taylor-ineq-k=2_no2}, the uniform upper bound on $K(\lambda, n, t_1)$ given in Lemma~\ref{lem:unif_upp_bd_on_K} in Appendix~\ref{appdx:bounds on K}, and the fact that $(\frac{t_1}{n}-\theta_1)^2\ge \delta^2 \theta_1^2$, \eqref{appi:final} follows because for $1\le t_1\le \lfloor n(1-\delta)\theta_1 \rfloor$, 
\begin{align}
t_1t_2 \ge n-1 \text{.}
\end{align}
Since the supremum in
\begin{align*}
\sup_{\theta_1\in \left[\frac{c\log n}{n}, \frac{1}{2} \right]} \hspace{-0.5mm}
(1-\delta)\theta_1 \hspace{-0.5mm} \left(\hspace{-0.5mm}\frac{9^\lambda \rme^\frac16 n^3}{2\pi(n-1)}\hspace{-0.5mm}\right)^{\frac12} \exp_\rme\hspace{-0.5mm} \left(\hspace{-0.5mm}-\frac{n}{2}(1+\lambda)  {\delta^2 \theta_1}\hspace{-0.5mm}\right)
\end{align*}
is attained at $\theta_1=\frac{c\log n}{n} $, it follows that \eqref{eqn:lem:achiv-k=2-S1} holds.
\end{IEEEproof}
\begin{lem}\label{lem:k=2_sum_S2_asympt}
Let $c \in (0, 1/(2\log \rme))$.
\begin{align}
\lim_{\delta\to 0} \limsup_{n \to \infty} \sup_{\theta_1\in \left[ \frac{c\log n}{n}, \frac{1}{2} \right]} S_2(\lambda, \theta_1, n, \delta) \le (1+\lambda)^{-\frac{1}{2}}\text{,} \label{eqn:lem:k=2-S2}
\end{align}
where $ S_2(\lambda, \theta_1, n, \delta)$ is defined in \eqref{eqn:k=2_sum_S2}.
\end{lem}
\begin{IEEEproof}
Assume that $n$ is a sufficiently large integer, let $\theta_1 \in \left[ \frac{c\log n}{n} , \frac{1}{2} \right] $ be given and define
\begin{align}
	\sigma_n=\sqrt{\frac{\theta_1(1-\theta_1) }{n(1+\lambda)(1-\delta)} }\text{.}
\end{align} 
We have
\begin{align}
&S_2(\lambda, \theta_1, n, \delta) \nonumber \\  
&\le \frac{M(2, \lambda, n, c, \delta) }{\sqrt{(1-\delta)^2 (1-(1+\delta)\theta_1) } } \sqrt{\frac{1-\theta_1 }{1+\lambda}}\label{bring1} \\ 
&\quad \times \sum_{t_1=\lceil n(1-\delta)\theta_1 \rceil }^{\lfloor n(1+\delta)\theta_1 \rfloor} \frac{1}{n} \frac{1}{\sqrt{ 2\pi} \sigma_n } \exp_\rme\left(-\frac{\left(\frac{t_1}{n}-\theta_1\right)^2}{2\sigma_n^2}\right)  \nonumber \\
&\le \frac{M(2, \lambda, n, c, \delta)}{\sqrt{(1-\delta)^3 (1+\lambda) } }  \label{bring2}\\ 
&\quad \times  \sum_{t_1=\lceil n(1-\delta)\theta_1 \rceil }^{\lfloor n(1+\delta)\theta_1 \rfloor} \frac{1}{n} \frac{1}{\sqrt{ 2\pi} \sigma_n } \exp_\rme\left(-\frac{\left(\frac{t_1}{n}-\theta_1\right)^2}{2\sigma_n^2}\right)\text{,}  \nonumber
\end{align}
where \eqref{bring1} is due to \eqref{taylor-ineq-k=2}, the bound on $K(\lambda, n, t_1) $ for the given range of $t_1$ (see Lemma~\ref{lem:limit_of_K} in Appendix~\ref{appdx:bounds on K}), and the fact that for $\lceil n(1-\delta)\theta_1 \rceil \le t_1 \le \lfloor n(1+\delta)\theta_1 \rfloor $,
\begin{align}
\sqrt{t_1 \left(1-t_1 \right)} \ge n\sqrt{(1-\delta)\theta_1(1-(1+\delta)\theta_1)}\text{,}
\end{align}
\eqref{bring2} follows because for $\theta_1 \in \left[ \frac{c\log n}{n} , \frac{1}{2} \right] $,
\begin{align}
	\sqrt{\frac{1-\theta_1}{1-(1+\delta)\theta_1}}&=\sqrt{1+\frac{\delta\theta_1}{1-(1+\delta)\theta_1} } \\ 
	&\le \frac{1}{\sqrt{1-\delta}}\text{.}
\end{align}
In light of Lemma~\ref{lem:limit_of_K} in Appendix~\ref{appdx:bounds on K},
\begin{align}
\lim_{n\to \infty} M(2, \lambda, n, c, \delta)=1\text{.}
\end{align}
Moreover, the Riemann sum in \eqref{bring2} can be upper bounded as
\begin{align}
\limsup_{n\to \infty }	\sum_{t_1=\lceil n(1-\delta)\theta_1 \rceil }^{\lfloor n(1+\delta)\theta_1 \rfloor} \frac{1}{n} \frac{1}{\sqrt{ 2\pi} \sigma_n } \exp_\rme\left(-\frac{(\frac{t_1}{n}-\theta_1)^2}{2\sigma_n^2}\right) \le 1\text{.}
\end{align}
It follows that \eqref{eqn:lem:k=2-S2} holds.
\end{IEEEproof}
\begin{lem}\label{lem:k=2_sum_S3_asympt}
Let $c\in (0, 1/(2\log \rme))$ and fix $\delta \in (0,1)$.
\begin{align}
\lim_{n \to \infty} \sup_{\theta_1\in \left[ \frac{c\log n}{n}, \frac{1}{2} \right]} S_3(\lambda, \theta_1, n, \delta)=0\text{,}
\end{align}
where $S_3(\lambda, \theta_1, n, \delta)$ is defined in \eqref{eqn:k=2_sum_S3}.
\end{lem}
\begin{IEEEproof}
The proof of this lemma is more involved than that of Lemma~\ref{lem:k=2_sum_S1_asympt}. To proceed, using Pinsker's inequality (e.g., \cite[Ex.~3.18]{csiszar-korner81}), namely
\begin{align}
d(\tau\|\theta)\ge 2(\tau-\theta)^2\text{,} \label{eqn:pinsker}
\end{align}
we first prove that 
\begin{align}
\lim_{n \to \infty} \sup_{\theta_1\in \left[ n^{-\frac{\beta}{2}}, \frac{1}{2} \right]} S_3(\lambda, \theta_1, n, \delta)=0\text{,} \label{eqn:proof:k=2-S3-no1}
\end{align}
where $\beta \in (0,1)$ is a fixed constant. Then, we show that
\begin{align}
\lim_{n \to \infty} \sup_{\theta_1\in \left[ \frac{c\log n}{n} , n^{-\frac{\beta}{2}} \right]} S_3(\lambda, \theta_1, n, \delta)=0\text{,} \label{eqn:proof:k=2-S3-no2}
\end{align}
with the help of Lemma~\ref{lem:bound on integral} in Appendix~\ref{appdx:bound on integral}.
Fix a constant $\beta \in (0,1)$, and assume that $n$ is a sufficiently large integer.

First, let $\theta_1 \in \left[ n^{-\frac{\beta}{2}}, \frac{1}{2} \right] $ be arbitrary and note that
\begin{align}
&S_3(\lambda, \theta_1, n, \delta)\nonumber \\
&\le 
\sum_{t_1=\lceil n(1+\delta)\theta_1 \rceil }^{n-1} 
\left(\frac{9^\lambda \rme^\frac16 n}{2\pi t_1t_2}\right)^\frac12 \exp_\rme\left({-2n(1+\lambda) \delta^2\theta_1^2}\right) \label{appf:from_t1_limits} \\
&\le
 \left(\frac{9^\lambda \rme^\frac16 n^3}{2\pi(n-1)}\right)^\frac12 \exp_\rme\left(-2(1+\lambda) \delta^2 n^{1-\beta}\right) \text{,} \label{appf:less_than_n_terms}
\end{align}
where \eqref{appf:from_t1_limits} follows from Lemma~\ref{lem:unif_upp_bd_on_K} in Appendix~\ref{appdx:bounds on K}, Pinsker's inequality as in \eqref{eqn:pinsker}, and the fact that $\left( \frac{t_1}{n}-\theta_1 \right)^2 \ge \delta^2 \theta_1^2$, \eqref{appf:less_than_n_terms} follows because $\theta_1 \ge n^{-\frac{\beta}{2}}$ and for $ \lceil n(1+\delta)\theta_1 \rceil \le t_1 \le n-1$, 
\begin{align}
t_1t_2 \ge (n-1)\text{.}
\end{align} 
Thus, \eqref{appf:less_than_n_terms} implies that 
\begin{align}
&\sup_{\theta_1\in \left[ n^{-\frac{\beta}{2}}, \frac{1}{2} \right]} S_3(\lambda, \theta_1, n, \delta) \nonumber \\  
&\qquad \le \left(\frac{9^\lambda \rme^\frac16 n^3}{2\pi(n-1)}\right)^\frac12 \exp_\rme\left(-2(1+\lambda) \delta^2 n^{1-\beta}\right) \text{.} 
\end{align}
Since $\beta <1$, 
\begin{align}
\lim_{n\to \infty}  \left(\frac{9^\lambda \rme^\frac16 n^3}{2\pi(n-1)}\right)^\frac12 \exp_\rme\left(-2(1+\lambda) \delta^2 n^{1-\beta}\right) =0\text{,}
\end{align}
and it follows that \eqref{eqn:proof:k=2-S3-no1} holds.

Second, let $\theta_1\in \left[ \frac{c\log n}{n} , n^{-\frac{\beta}{2}} \right]$ be arbitrary and fix some constant $ \kappa \in \left(0, \frac{1}{2}\right)$. Further, separate $S_3(\lambda, \theta_1, n, \delta)$ into two sums as follows
\begin{align}
S_3(\lambda, \theta_1, n, \delta)=\widetilde{S}_3^1(\kappa, \lambda, \theta_1, n, \delta)+\widetilde{S}_3^2(\kappa, \lambda, \theta_1, n, \delta)\text{,}
\end{align}
where
\begin{align}
\widetilde{S}_3^1(\kappa, \lambda, \theta_1, n, \delta)
&=
\sum_{t_1=\lceil n\kappa \rceil }^{n-1}  
\left(\frac{n}{2\pi t_1t_2}\right)^\frac12 \\
& \hspace{-6mm} \times \exp_\rme\left(-n(1+\lambda) d\left(\tfrac{t_1}{n}\|\theta_1\right)\right) K(\lambda, n, t_1)\text{,} \nonumber  \\
\widetilde{S}_3^2(\kappa, \lambda, \theta_1, n, \delta) 
&= 
\sum_{t_1=\lceil n(1+\delta)\theta_1 \rceil }^{\lfloor n\kappa \rfloor} \left(\frac{n}{2\pi t_1t_2}\right)^\frac12  \\ 
&\hspace{-6mm} \times \exp_\rme\left(-n(1+\lambda) d\left(\tfrac{t_1}{n}\|\theta_1\right)\right) K(\lambda, n, t_1)\text{.} \nonumber
\end{align}
Regarding $\widetilde{S}_3^1(\kappa, \lambda, \theta_1, n, \delta)$, we have
\begin{align}
\widetilde{S}_3^1(\kappa, \lambda, \theta_1, n, \delta) 
&\le  \sum_{t_1=\lceil n\kappa \rceil }^{n-1}  \sqrt{\frac{n}{2\pi t_1t_2}} \label{appe:from_pins} \\
&\hspace{-6mm} \times \exp_\rme\left(-2n(1+\lambda) \left( \tfrac{t_1}{n}-\theta_1\right)^2\right)3^\lambda\rme^{\frac{1}{12}} \nonumber \\
&\hspace{-19mm} \le 
\frac{n}{\sqrt{2\pi\kappa}} \exp_\rme\left(-2n(1+\lambda) \big( \kappa - n^{-\frac{\beta}{2}} \big)^2\right)3^\lambda \rme^{\frac{1}{12}}\text{,} \label{appe:last_ineq} 
\end{align}
where \eqref{appe:from_pins} follows from Lemma~\ref{lem:unif_upp_bd_on_K} in Appendix~\ref{appdx:bounds on K} and \eqref{eqn:pinsker}, \eqref{appe:last_ineq} follows because $\frac{t_1}{n}-\theta_1 \ge \kappa-n^{-\frac{\beta}{2}} $ and $\sqrt{t_1t_2}\ge \sqrt{n\kappa}$ for $\lceil n\kappa \rceil \le t_1 \le n-1 $ and $c\log n /n\le \theta_1 \le n^{-\frac{\beta}{2}}$. Hence,
\begin{align}
& \sup_{\theta_1 \in \left[ \frac{c\log n}{n} , n^{-\frac{\beta}{2}} \right] }\widetilde{S}_3^1(\kappa, \lambda, \theta_1, n, \delta) \nonumber \\
&\quad \le \frac{n}{\sqrt{2\pi\kappa}} \exp_\rme\left(-2n(1+\lambda) \big( \kappa - n^{-\frac{\beta}{2}} \big)^2\right)3^\lambda \rme^{\frac{1}{12}}\text{,} \label{appe:sup_last_ineq} 
\end{align}
and 
\begin{align}
	\lim_{n\to \infty} \sup_{\theta_1 \in \left[ \frac{c\log n}{n} , n^{-\frac{\beta}{2}} \right] }\widetilde{S}_3^1(\kappa, \lambda, \theta_1, n, \delta) = 0\text{.} \label{limit_of_stilde1}
\end{align}
Regarding $\widetilde{S}_3^2(\kappa, \lambda, \theta_1, n, \delta)$, we have
\begin{align}
&\frac{\sqrt{2\pi}}{3^\lambda \rme^{\frac{1}{12}}} \widetilde{S}_3^2(\kappa, \lambda, \theta_1, n, \delta) \nonumber \\ 
&\ \le \sum_{t_1=\lceil n(1+\delta)\theta_1 \rceil }^{\lfloor n\kappa \rfloor }
\frac{\sqrt{n} }{\sqrt{t_1t_2}} \exp_\rme\left(-n(1+\lambda) d\left(\tfrac{t_1}{n}\|\theta_1\right)\right)\text{,} \label{appd:from_lem_K} 
\end{align} 
where \eqref{appd:from_lem_K} follows from Lemma~\ref{lem:unif_upp_bd_on_K} in Appendix~\ref{appdx:bounds on K}. Let $\theta_1^\ast \in \big[ \frac{c\log n}{n} , n^{-\frac{\beta}{2}} \big]$ be the maximizer of the right side in \eqref{appd:from_lem_K}. 

Note that 
\begin{align}
	&\limsup_{n\to \infty}   \sum_{t_1=\lceil n(1+\delta)\theta_1^\ast \rceil }^{\lfloor n\kappa \rfloor } 
\frac{\sqrt{n} }{\sqrt{t_1t_2}} \rme^{-n(1+\lambda) d\left(\frac{t_1}{n}\|\theta_1^\ast\right)} \nonumber \\  
&\quad \le \limsup_{n\to \infty}  	\int_{(1+\delta)\theta_1^\ast}^{\kappa } \frac{\sqrt{n}}{\sqrt{\tau(1-\tau)}}\rme^{-n(1+\lambda)d(\tau\|\theta_1^\ast)}\rmd \tau \label{araf:riemann} \\
&\quad \le  \limsup_{n\to \infty}  \frac{(1+\lambda)^{-1}\ln^{-1}(1+\delta)}{\sqrt{n(1+\delta)\theta_1^\ast(1-(1+\delta)\theta_1^\ast)} } \label{araf:integral-lem} \\
&\quad \le  \limsup_{n\to \infty}   \frac{(1+\lambda)^{-1} \ln^{-1}(1+\delta)}{\sqrt{(1+\delta)c\log n \big(1-(1+\delta) n^{-\frac{\beta}{2}} \big)} } \\ 
&\quad = 0\text{,}
\end{align} 
where \eqref{araf:riemann} follows after noticing that for any $\theta\in \big[\frac{c\log n}{n} , n^{-\frac{\beta}{2}} \big]$, the function
\begin{align}
	g_\theta(x)=\frac{1}{\sqrt{x(1-x)}}\rme^{-n(1+\lambda)d(x\| \theta) }
\end{align}
is a decreasing function in $x\in ((1+\delta)\theta,\kappa)  $ and therefore the corresponding Riemann sum in the left side  of \eqref{araf:riemann} can be upper bounded by the integral in its right side and \eqref{araf:integral-lem} follows from Lemma~\ref{lem:bound on integral} in Appendix~\ref{appdx:bound on integral}. Hence, 
\begin{align}
	\lim_{n\to \infty} \sup_{\theta_1 \in \big[ \frac{c\log n}{n} , n^{-\frac{\beta}{2}} \big] }\widetilde{S}_3^2(\kappa, \lambda, \theta_1, n, \delta) = 0\text{.} \label{limit_of_stilde2}
\end{align}
As a result of \eqref{limit_of_stilde1} and \eqref{limit_of_stilde2}, \eqref{eqn:proof:k=2-S3-no2} holds. The desired result follows since we have established \eqref{eqn:proof:k=2-S3-no1} and \eqref{eqn:proof:k=2-S3-no2}.

\end{IEEEproof} 
%%%%%%%%%%%%%%%%%%
\section{Upper Bound for the Integral in \eqref{araf:riemann}} \label{appdx:bound on integral}
\begin{lem}\label{lem:bound on integral}
Let $c\in (0,1/(2\log \rme)) $ and $\lambda\in (0,\infty) $. Fix $\beta\in(0,1) $, $\delta\in(0,1) $ and $\kappa\in(0,1/2) $. For any $\theta_1\in \left[ \frac{c\log n}{n}, n^{-\frac{\beta}{2}} \right] $
\begin{align}
	&\int_{(1+\delta)\theta_1}^{\kappa } \frac{n(1+\lambda)}{\sqrt{\tau(1-\tau)}}\rme^{-n(1+\lambda)d(\tau\|\theta_1)}\rmd \tau 
	\nonumber \\ &\qquad \qquad \quad \le \frac{\ln^{-1}(1+\delta)}{\sqrt{(1+\delta)\theta_1(1-(1+\delta)\theta_1)} }\text{.}
\end{align}
\end{lem}
\begin{IEEEproof}
Abbreviate 
\begin{align}
	a_n&=n(1+\lambda)\text{,} \\
	\varphi(\tau)&=\frac{1}{\sqrt{\tau(1-\tau)}}\text{,} \\ 
	\phi(\tau) &= d(\tau\|\theta_1)\text{.}
\end{align}
Applying integration by parts yields
\begin{align}
	&\int  a_n \varphi(\tau)\rme^{-a_n\phi(\tau)}\rmd \tau \nonumber \\
	&\ \ \, = -\frac{\varphi(\tau)}{\phi'(\tau)}\rme^{-a_n\phi(\tau)}+ \int \rme^{-a_n\phi(\tau)} \frac{\rmd}{\rmd \tau} \left(\frac{\varphi(\tau)}{\phi'(\tau)} \right) \rmd \tau \text{.} 
\end{align}
For $\tau \in \left[(1+\delta)\theta_1, \kappa \right] $, we have
\begin{align}
	\frac{\rmd}{\rmd \tau}\left(\frac{\varphi(\tau)}{\phi'(\tau)}\right) \le 0\text{,}
\end{align}
because $\varphi(\tau)$ is a decreasing function and $\phi(\tau) $ is an increasing convex function for the given range of $\tau $. Hence, we see that
\begin{align}
\int_{(1+\delta)\theta_1}^{\kappa}  a_n \varphi(\tau)\rme^{-a_n\phi(\tau)}\rmd \tau &\le 
\frac{\varphi(\tau)}{\phi'(\tau)}\rme^{-a_n\phi(\tau)}\bigg|_{\tau=\kappa}^{(1+\delta)\theta_1}\\
& \le 
\frac{\varphi((1+\delta)\theta_1)}{\ln(1+\delta) } \label{cute2}\text{,}
\end{align}
where \eqref{cute2} follows because $\kappa\le 1/2 $ implies
\begin{align}
	\frac{\varphi(\tau)}{\phi'(\tau)}\rme^{-a_n\phi(\tau)}\bigg|_{\tau=\kappa} \ge 0\text{,}
\end{align}
and
\begin{align}
\phi'(\tau)\rme^{a_n\phi(\tau)}\bigg|_{\tau=(1+\delta)\theta_1} \ge \ln (1+\delta)\text{.}
\end{align}	
\end{IEEEproof}
%%%%%%%%%%%%%%
\section{Lemmas for the Proof in Section~\ref{sec:subsec:ach-gen}} \label{appdx:achievability_k>2}
In the proofs of Lemmas~\ref{lem:generela_sum_S1} and~\ref{lem:generela_sum_S2}, we use the following bound: for $\theta_k \ge 1/k $ and $\delta\in (0, 1/(k-1)) $, 
\begin{align}
&|\tau_i-\theta_i|\le \delta \theta_i \text{ for } i=1, \hdots, k-1 \implies  \nonumber \\ & D(\boldtau\| \boldtheta) \ge 
\frac{1}{2}(\boldtau'-\boldtheta')^{T}\bfJ(\boldtheta, P_{Y|V})(\boldtau'-\boldtheta')(1-(k-1)\delta)\text{,} \label{Taylor-gen-bound}
\end{align}
where $\bfJ(\boldtheta, P_{Y|V} ) $ denotes the Fisher information matrix, and 
\begin{align}
\boldtau'&=(\tau_1, \hdots, \tau_{k-1})\text{,} \\
\boldtheta'&=(\theta_1, \hdots, \theta_{k-1})\text{.}
\end{align}
To show \eqref{Taylor-gen-bound}, we rely on Taylor's theorem:
\begin{align}
D(\boldtau\|\boldtheta)=\sum_{i=1}^{k}\left( \frac{(\tau_i-\theta_i)^2}{2\theta_i}-\frac{(\tau_i-\theta_i)^3}{6\alpha_i^2} \right)\text{,} \label{Taylor-gen-eql}
\end{align}
for some $\boldalpha=(\alpha_1, \hdots, \alpha_k)\in \simplex^{k-1} $ such that $\alpha_i$ lies between $\tau_i $ and $\theta_i $.
\begin{lem} \label{lem:generela_sum_S1}
The function defined in \eqref{eqn:def:general-S1} satisfies
\begin{align}
\lim_{\delta \to 0}\limsup_{n\to \infty} \sup_{\substack{\boldtheta \in \cR_0 \\ \theta_k \ge 1/k}} S_1(k, \lambda, \boldtheta,n,\delta)  \le (1+\lambda)^{-\frac{k-1}{2}}\text{.} \label{eqn:lem:general-S1}
\end{align}
\end{lem}
\begin{IEEEproof}
Assume that $n$ is a sufficiently large integer, and let $\boldtheta\in \cR_0$ with $\theta_k \ge 1/k$ be given. Define
\begin{align}
\mathbold{\Sigma}_n&=\frac{\bfJ^{-1}(\boldtheta, P_{Y|V})}{n(1+\lambda)(1-(k-1)\delta)}\text{.}
\end{align}
We invoke \eqref{Taylor-gen-bound} with
\begin{align}
	\boldtau'\leftarrow \big( \tfrac{t_1}{n}, \hdots, \tfrac{t_{k-1}}{n} \big)\text{.}
\end{align}
Hence,
\begin{align}
	S_1(k, \lambda, \boldtheta, n, \delta) &\le \frac{M(k, \lambda, n, c, \delta)}{(1-\delta)^{\frac{k-1}{2}}(1-(k-1)\delta)^{\frac{k-1}{2}}} \label{bulasik-mak}    \\ 
	& \hspace{-20mm} \times\sqrt{\frac{\theta_k (1+\lambda)^{1-k} }{(1+\delta) \theta_k-\delta}}  \sum_{\substack{ \boldt \colon \boldt \in \cN_\delta^\boldtheta  \\ t_i\ge 1 \ \forall i }} \frac{\rme^{-\frac{1}{2}(\boldtau'-\boldtheta')^T \mathbold{\Sigma}_n^{-1}(\boldtau'-\boldtheta') }}{n^{k-1}(2\pi)^{\frac{k-1}{2}} |\mathbold{\Sigma}_n|^\frac{1}{2}} \nonumber \\
	&\le \frac{(1+\lambda)^{\frac{k-1}{2}} M(k, \lambda, n, c, \delta)}{(1-\delta)^{\frac{k-1}{2}}(1-(k-1)\delta)^{\frac{k}{2}}}\label{bulasik-mak-durdu} \\
	& \hspace{5mm} \times \sum_{\substack{ \boldt \colon \boldt \in \cN_\delta^\boldtheta   \\ t_i\ge 1 \ \forall i }} \frac{\rme^{-\frac{1}{2}(\boldtau'-\boldtheta')^T \mathbold{\Sigma}_n^{-1}(\boldtau'-\boldtheta') }}{n^{k-1}(2\pi)^{\frac{k-1}{2}} |\mathbold{\Sigma}_n|^\frac{1}{2}}\text{,}  \nonumber
\end{align}
where \eqref{bulasik-mak} is due to \eqref{Taylor-gen-bound}, the bound on $K(k,\lambda, n, \boldt)$ when $\boldt \in \cN_\delta^\boldtheta$ (see Lemma~\ref{lem:limit_of_K} in Appendix~\ref{appdx:bounds on K}), and the fact that for $\boldt \in \cN_\delta^\boldtheta$, 
\begin{align}
	\prod_{i=1}^k  t_i^{\frac12} \ge n^{\frac k2} (1-\delta)^{\frac{k-1}{2}} \sqrt{\theta_1 \cdots \theta_{k-1} }\sqrt{(1+\delta)\theta_k-\delta}\text{,}  
\end{align}  
\eqref{bulasik-mak-durdu} follows because $\theta_k \ge 1/k $ implies
\begin{align}
	\frac{\theta_k}{(1+\delta)\theta_k-\delta} &\le \frac{1}{1-(k-1)\delta} \text{.}
\end{align}
In light of Lemma~\ref{lem:limit_of_K} in Appendix~\ref{appdx:bounds on K},
\begin{align}
	\lim_{n\to \infty} M(k, \lambda, n, c, \delta)=1\text{.}
\end{align}
Since the multi-variable Riemann sum in \eqref{bulasik-mak-durdu} can be upper bounded as
\begin{align}
\limsup_{n\to \infty }	\sum_{\substack{ \boldt \colon \boldt \in \cN_\delta^\boldtheta \\ t_i\ge 1 \ \forall i}} \frac{\rme^{-\frac{1}{2}(\boldtau'-\boldtheta')^T \mathbold{\Sigma}_n^{-1}(\boldtau'-\boldtheta') }}{n^{k-1}(2\pi)^{\frac{k-1}{2}} |\mathbold{\Sigma}_n|^\frac{1}{2}} \le 1\text{,}
\end{align}
we can conclude that \eqref{eqn:lem:general-S1} holds.
\end{IEEEproof}
\begin{lem}\label{lem:generela_sum_S2}
The function defined in \eqref{eqn:def:general-S2} satisfies
\begin{align}
\lim_{n \to \infty}\sup_{\substack{\boldtheta \in \cR_0 \\ \theta_k \ge 1/k}} S_2(k, \lambda, \boldtheta,n,\delta) =0\text{.} \label{eqn:lem:general-S2}
\end{align}	
\end{lem}
\begin{IEEEproof}
Assume that $n$ is a sufficiently large integer, and let $\boldtheta \in \cR_0$ with $\theta_k \ge 1/k$ be given. 
Recall the definition of $\cN_\delta^\boldtheta$ in \eqref{eqn:defn:N-delta}, and note that if 
\begin{align}
\boldt  \not \in  \cN^\boldtheta_\delta\text{,}
\end{align}
then there must exist $i \in \{1,\hdots, k-1\}$ such that 
\begin{align}
t_i \not \in  I_{\delta, \theta_i, n}=\left[ \lceil n(1-\delta)\theta_i  \rceil , \lfloor n(1+\delta)\theta_i \rfloor  \right]\text{.} 
\end{align}
Moreover, by symmetry, we can write
\begin{align}
& S_2(k, \lambda, \boldtheta,n,\delta) \nonumber \\
&=\sum_{\substack{ 1\le t_1 \le n-k+1 \\ t_i\ge 1 \ \forall i \\ t_1 \not \in I_{\delta, \theta_1, n} }} \sum_{\substack{t_2, \hdots, t_k \\ t_2+\cdots+t_k=n-t_1}} \frac{K(k,\lambda, n, \boldt)}{(2\pi)^{\frac{k-1}{2}} } \left(\frac{n}{\prod_{i=1}^{k}t_i}\right)^\frac12 \nonumber \\ 
&\qquad \qquad \times \exp_\rme(-n(1+\lambda)D(\widehat{P}_{y^n}\| P_{Y|V=\boldtheta}))  \\
&\le 
\sum_{\substack{ 1\le t_1 \le n-(k-1) \\ t_i\ge 1 \ \forall i \\ t_1 \not \in I_{\delta, \theta_1, n} }} \sum_{\substack{t_2, \hdots, t_k \\ t_2+\cdots+t_k=n-t_1}} \frac{M(k, \lambda)}{(2\pi)^{\frac{k-1}{2}} } \left(\frac{n}{\prod_{i=1}^{k}t_i}\right)^\frac12 \nonumber \\
& \qquad \qquad \times \exp_\rme(-n(1+\lambda)D(\widehat{P}_{y^n}\| P_{Y|V=\boldtheta})) \label{aqtz}\\
&=
\sum_{\substack{ 1\le t_1 \le n-(k-1) \\ t_i\ge 1 \ \forall i \\ t_1 \not \in I_{\delta, \theta_1, n} }} 
\left(\frac{ (2\pi)^{-1} n  }{ t_1(n-t_1)}\right)^\frac12 \exp_\rme(-n(1+\lambda) d(\tfrac{t_1}{n}\|\theta_1))  \nonumber  \\ 
&\qquad \qquad \times M(k, \lambda) \mfrakT(k-1, \lambda, \boldtheta', n-t_1)\text{,}  
  \label{aqty}
\end{align}
where (\ref{aqtz}) is due to the uniform upper bound on $K(k, \lambda, n, \boldt) $ in Lemma~\ref{lem:unif_upp_bd_on_K}, in (\ref{aqty}), $\boldtheta'= \left( \frac{\theta_2}{1-\theta_1}, \cdots, \frac{\theta_k}{1-\theta_1} \right) $ and the function denoted by $\mfrakT(k, \lambda, \boldtheta, n) $ is defined in \eqref{eqn:def:mfrakT-general}. By invoking Lemma~\ref{lem:uniform bound on the sum} in Appendix~\ref{appdx:uniform upper bound on the sum}, we see that $\mfrakT(k-1, \lambda, \boldtheta', n-t_1) $ can be upper bounded by a constant depending only on $\lambda$ and $k$. On the other hand, the sum without the factor $\mfrakT $ vanishes as $ n\to \infty $ (see Lemmas~\ref{lem:k=2_sum_S1_asympt} and~\ref{lem:k=2_sum_S3_asympt}). Therefore, \eqref{eqn:lem:general-S2} follows.  
\end{IEEEproof} 
%%%%%%%%%%%%%%%%%%%%%%%%
\section{Jeffreys' Mixture is not Minimax}\label{appdx:jeffreys' mixture is not minimax}
The fact that Jeffreys' prior is capacity achieving (or least favorable) follows from the converse proof of Theorem~\ref{thm:Asymptotic Behavior of Minimax Renyi Redundancy}. Therefore, Jeffreys' mixture is maximin for R\'enyi redundancy. Parallel to the results in \cite{XieBarron1997} and \cite{XieBarron2000}, Lemma~\ref{lem:jefreys mixture is not minimax} below proves that Jeffreys' mixture is \emph{not} minimax. 
\begin{lem} \label{lem:jefreys mixture is not minimax}
For any $l \in \{1,\hdots, k-1 \}$, 
\begin{align}
&\liminf_{n\to \infty }\left\{  \sup_{\boldtheta} D_{1+\lambda}(P_{Y^n|V=\boldtheta} \| Q_{Y^n}^\ast ) - \frac{k-1}{2} \log \frac{n}{2\pi} \right\} \nonumber \\ 
&\qquad  \ge \log \frac{\opGamma^k(1/2) }{\opGamma(k/2) }-\frac{k-1}{2\lambda} \log(1+\lambda) \label{eqn:lem:jeffreys mixture is not minimax} \\ 
&\qquad \qquad
+\frac{k-l}{2} \left( \log 2 + \frac{\log (1+\lambda) }{\lambda } \right)\text{,} \nonumber
\end{align}
where the supremization is over all $\boldtheta\in \simplex^{k-1}$ that are on the face of the simplex so that at most $l $ of its components are known to be non-zero. 
\end{lem}
Note that the third term in the right side of \eqref{eqn:lem:jeffreys mixture is not minimax} interpolates the extra constants $\frac{k-l}{2} \log(2\rme)  $ when $\lambda=0$  and $\frac{k-l}{2} \log2  $ when $\lambda=\infty $, shown in \cite{XieBarron1997} and \cite{XieBarron2000}, respectively.
\begin{IEEEproof}
		Assuming without loss of generality that the last $k-l$ entries of $\boldtheta $ are equal to zero simplifies the notation. Otherwise, the proof remains identical. Define
		\begin{align}
			\bar{\boldtheta}&=(\theta_1, \hdots, \theta_l ) \in \simplex^{l-1} \text{,}\\
		L(k,l,n)	&= \frac{\left( 1+\frac{k}{2n} \right)^{n+\frac{k-1}{2} } }{\left( 1+ \frac{l}{2n} \right)^{n+\frac{l-1}{2} } }\frac{2-\rme^{\frac{1}{12n+6k} } }{\rme^{\frac{k-l}{2}+\frac{1}{12n+6l} } }\text{,}
		\end{align}
where $\theta_i$ denotes the $i$-th entry of $\boldtheta$. Note that
\begin{align}
&D_{1+\lambda}(P_{Y^n|V=\boldtheta} \| Q_{Y^n}^\ast )\nonumber \\ &= D_{1+\lambda}(P_{Y^n|V=\bar{\boldtheta}} \| Q_{Y^n}^{\ast(l-1)} )  +\log \frac{\opGamma(\frac l2)\opGamma(n+\frac k2) }{\opGamma(\frac k2) \opGamma(n+\frac l2)} \label{trytoforget} \\ 
&\ge 
D_{1+\lambda}(P_{Y^n|V=\bar{\boldtheta}} \| Q_{Y^n}^{\ast(l-1)} ) +\log \frac{\opGamma(\frac{l}{2})}{\opGamma(\frac{k}{2})} \label{trytoforget2}\\ 
&\qquad +\frac{k-l}{2}\log n + \log L(k, l, n)  \text{,} \nonumber
\end{align}
where $Q_{Y^n}^{\ast(l-1)}$ denotes the Jeffreys' mixture when the underlying parameter space is the $(l-1)$-dimensional simplex, \eqref{trytoforget} follows from the fact that
\begin{align}
&\frac{\opD_k(1/2, \hdots, 1/2) }{\opD_k(t_1+1/2, \hdots, t_l+1/2, 1/2, \hdots, 1/2) } \nonumber \\ 
&= \frac{\opD_l(1/2, \hdots, 1/2)}{\opD_l(t_1+1/2, \hdots, t_l+1/2 ) }\frac{\opGamma \left( l/2 \right) \opGamma(n+k/2)}{\opGamma \left(k/2\right) \opGamma\left(n+l/2 \right) }\text{,}
\end{align}
and \eqref{trytoforget2} follows from Stirling's approximation which can be seen in \eqref{eqn:Stirling_Approximation_for_Gamma_Function}. Since
\begin{align}
&\sup_{\boldtheta } D_{1+\lambda}(P_{Y^n|V=\boldtheta} \| Q_{Y^n}^\ast ) \nonumber \\ 
&\ge \sup_{\bar{\boldtheta}\in \simplex^{l-1} } D_{1+\lambda}(P_{Y^n|V=\bar{\boldtheta}} \| Q_{Y^n}^{\ast(l-1) } ) +\log \frac{\opGamma(l/2) }{\opGamma(k/2)}   \\ \nonumber 
&\qquad \qquad +\frac{k-l}{2}\log n + \log L(k, l, n) \label{son-label} \\ 
&\ge \inf_{Q_{Y^n}} \sup_{\bar{\boldtheta}\in \simplex^{l-1} }  D_{1+\lambda}(P_{Y^n|V=\bar{\boldtheta}} \| Q_{Y^n} ) +\log \frac{\opGamma(l/2) }{\opGamma(k/2)} \\ \nonumber 
&\qquad \qquad  +\frac{k-l}{2}\log n + \log L(k, l, n)\text{,}
\end{align}
where the supremization in the left side of \eqref{son-label} is over all $\boldtheta$ whose last $k-l$ entries are zero, the converse result in Section~\ref{sec:converse} with $k\leftarrow l$, and the fact that 
\begin{align}
	\lim_{n \to \infty} L(k,l,n) =1\text{,}
\end{align}
along with routine algebraic manipulations yield the desired result in \eqref{eqn:lem:jeffreys mixture is not minimax}.
\end{IEEEproof}

\end{appendices}

\section*{Acknowledgments}
This work has been supported by ARO-MURI contract number W911NF-15-1-0479 and in part by the Center for Science of Information, an NSF Science and Technology Center under Grant CCF-0939370.

\bibliographystyle{IEEEtran}
\bibliography{Minimax-Renyi-Revised_Submitted_v2.5.bib}

\begin{thebibliography}{10}
\providecommand{\url}[1]{#1}
\csname url@samestyle\endcsname
\providecommand{\newblock}{\relax}
\providecommand{\bibinfo}[2]{#2}
\providecommand{\BIBentrySTDinterwordspacing}{\spaceskip=0pt\relax}
\providecommand{\BIBentryALTinterwordstretchfactor}{4}
\providecommand{\BIBentryALTinterwordspacing}{\spaceskip=\fontdimen2\font plus
\BIBentryALTinterwordstretchfactor\fontdimen3\font minus
  \fontdimen4\font\relax}
\providecommand{\BIBforeignlanguage}[2]{{%
\expandafter\ifx\csname l@#1\endcsname\relax
\typeout{** WARNING: IEEEtran.bst: No hyphenation pattern has been}%
\typeout{** loaded for the language `#1'. Using the pattern for}%
\typeout{** the default language instead.}%
\else
\language=\csname l@#1\endcsname
\fi
#2}}
\providecommand{\BIBdecl}{\relax}
\BIBdecl

\bibitem{Yagli2017ISIT}
S.~Yagli, Y.~Altu\u{g}, and S.~Verd\'u, ``Minimax {R\'enyi} redundancy,'' in
  \emph{2017 IEEE International Symposium on Information Theory (ISIT)}, June
  2017, pp. 2980--2984.

\bibitem{CoverThomas}
T.~M. Cover and J.~A. Thomas, \emph{Elements of Information Theory},
  2nd~ed.\hskip 1em plus 0.5em minus 0.4em\relax Hoboken, NJ, USA: Wiley, 2006.

\bibitem{KontoyiannisVerdu2014}
I.~Kontoyiannis and S.~Verd{\'u}, ``Optimal lossless data compression:
  non-asymptotics and asymptotics,'' \emph{IEEE Transactions on Information
  Theory}, vol.~60, no.~2, pp. 777--795, Feb. 2014.

\bibitem{Gallagher1976}
R.~G. Gallager, ``Source coding with side information and universal coding,''
  September 1976, unpublished manuscript. Available from:
  \url{http://web.mit.edu/gallager/www/papers/paper5.pdf}.

\bibitem{Ryabko1979}
B.~Y. Ryabko, ``Coding of a source with unknown but ordered probabilities,''
  \emph{Problems of Information Transmission}, vol.~15, no.~2, pp. 134--138,
  Oct. 1979.

\bibitem{XieBarron2000}
Q.~Xie and A.~R. Barron, ``Asymptotic minimax regret for data compression,
  gambling, and prediction,'' \emph{IEEE Transactions on Information Theory},
  vol.~46, no.~2, pp. 431--445, Mar. 2000.

\bibitem{Shtarkov87}
Y.~M. Shtarkov, ``Universal sequential coding of single messages,''
  \emph{Problemy Peredachi Informatsii}, vol.~23, no.~3, pp. 3--17, Jul.--Sep.
  1987.

\bibitem{ForsterWarmuth2002}
J.~Forster and M.~K. Warmuth, ``Relative expected instantaneous loss bounds,''
  \emph{Journal of Computer and System Sciences}, vol.~64, no.~1, pp. 76--102,
  Feb. 2002.

\bibitem{Dramota04}
M.~Drmota and W.~Szpankowski, ``Precise minimax redundancy and regret,''
  \emph{IEEE Transactions on Information Theory}, vol.~50, no.~11, pp.
  2686--2707, Nov. 2004.

\bibitem{LiangBarron2004}
F.~Liang and A.~R. Barron, ``Exact minimax strategies for predictive density
  estimation, data compression, and model selection,'' \emph{IEEE Transactions
  on Information Theory}, vol.~50, no.~11, pp. 2708--2726, Nov. 2004.

\bibitem{Pratt64}
J.~W. Pratt, ``Risk aversion in the small and in the large,''
  \emph{Econometrica: Journal of the Econometric Society}, vol.~32, no. 1--2,
  pp. 122--136, Jan.--Apr. 1964.

\bibitem{Arrow65}
K.~J. Arrow, \emph{Aspects of the Theory of Risk-Bearing}.\hskip 1em plus 0.5em
  minus 0.4em\relax Helsinki, Finland: Yrj{\"o} Jahnssonin S{\"a}{\"a}ti{\"o},
  1965.

\bibitem{Ross81}
S.~A. Ross, ``Some stronger measures of risk aversion in the small and the
  large with applications,'' \emph{Econometrica: Journal of the Econometric
  Society}, vol.~49, no.~3, pp. 621--638, May 1981.

\bibitem{Campbell1965}
L.~L. Campbell, ``A coding theorem and {R{\'e}nyi's} entropy,''
  \emph{Information and Control}, vol.~8, no.~4, pp. 423--429, Aug. 1965.

\bibitem{Sundaresan2007}
R.~Sundaresan, ``Guessing under source uncertainty,'' \emph{IEEE Transactions
  on Information Theory}, vol.~53, no.~1, pp. 269--287, Jan. 2007.

\bibitem{sibson1969information}
R.~Sibson, ``Information radius,'' \emph{Zeitschrift f{\"u}r
  Wahrscheinlichkeitstheorie und verwandte Gebiete}, vol.~14, no.~2, pp.
  149--160, 1969.

\bibitem{Verdu2015}
S.~Verd{\'u}, ``$\alpha$-mutual information,'' in \emph{2015 Information Theory
  and Applications Workshop}, San Diego, Feb. 2015, pp. 1--6.

\bibitem{Csiszar95}
I.~Csisz{\'a}r, ``Generalized cutoff rates and {R{\'e}nyi's} information
  measures,'' \emph{IEEE Transactions on Information Theory}, vol.~41, no.~1,
  pp. 26--34, Jan. 1995.

\bibitem{XieBarron1997}
Q.~Xie and A.~R. Barron, ``Minimax redundancy for the class of memoryless
  sources,'' \emph{IEEE Transactions on Information Theory}, vol.~43, no.~2,
  pp. 646--657, Mar. 1997.

\bibitem{Davisson81}
L.~D. Davisson, R.~J. McEliece, M.~B. Pursley, and M.~S. Wallace, ``Efficient
  universal noiseless source codes,'' \emph{IEEE Transactions on Information
  Theory}, vol.~27, no.~3, pp. 269--279, May 1981.

\bibitem{Gyorfi94}
L.~Gy\"orfi, I.~P\'ali, and E.~C. van~der Meulen, ``There is no universal
  source code for an infinite source alphabet,'' \emph{IEEE Transactions on
  Information Theory}, vol.~40, no.~1, pp. 267--271, Jan. 1994.

\bibitem{KriTro81}
R.~E. Krichevsky and V.~K. Trofimov, ``The performance of universal encoding,''
  \emph{IEEE Transactions on Information Theory}, vol.~27, no.~2, pp. 199--207,
  Mar. 1981.

\bibitem{Rissanen84}
J.~Rissanen, ``Universal coding, information, prediction, and estimation,''
  \emph{IEEE Transactions on Information Theory}, vol.~30, no.~4, pp. 629--636,
  Jul. 1984.

\bibitem{Rissanen86}
{J. Risannen}, ``Stochastic complexity and modeling,'' \emph{The Annals of
  Statistics}, vol.~14, no.~3, pp. 1080--1100, Sep. 1986.

\bibitem{Merhav11}
N.~Merhav, ``On optimum strategies for minimizing the exponential moments of a
  loss function,'' \emph{Communications in Information and Systems}, vol.~11,
  no.~4, pp. 343--368, 2011.

\bibitem{Hayashi17}
M.~Hayashi, ``Universal channel coding for general output alphabet,'' 2015,
  [Online] Available: https://arxiv.org/abs/1502.02218.

\bibitem{ClarkeBarron}
B.~S. Clarke and A.~R. Barron, ``Information-theoretic asymptotics of {Bayes}
  methods,'' \emph{IEEE Transactions on Information Theory}, vol.~36, no.~3,
  pp. 453--471, May 1990.

\bibitem{Jeffreys46}
H.~Jeffreys, ``An invariant form for the prior probability in estimation
  problems,'' in \emph{Proceedings of the Royal Society of London Series A:
  Mathematical, Physical and Engineering Sciences}, vol. 186, no. 1007, Sep.
  1946, pp. 453--461.

\bibitem{Erven2014}
T.~van Erven and P.~Harremo\"es, ``R{\'e}nyi divergence and {Kullback-Leibler}
  divergence,'' \emph{IEEE Transactions on Information Theory}, vol.~60, no.~7,
  pp. 3797--3820, Jul. 2014.

\bibitem{arimoto75information}
S.~Arimoto, ``Information measures and capacity of order {$\alpha $} for
  discrete memoryless channels,'' in \emph{Topics in Information Theory, Proc.
  Coll. Math. Soc. J\'anos Bolyai}.\hskip 1em plus 0.5em minus 0.4em\relax
  Keszthely, Hungary: Bolyai, 1975, pp. 41--52.

\bibitem{Robbins}
H.~Robbins, ``A remark on {Stirling's} formula,'' \emph{The American
  Mathematical Monthly}, vol.~62, no.~1, pp. 26--29, Jan. 1955.

\bibitem{Whittaker}
E.~T. Whittaker and G.~N. Watson, \emph{A Course of Modern Analysis},
  4th~ed.\hskip 1em plus 0.5em minus 0.4em\relax Cambridge, U.K.: Cambridge
  University Press, 1963.

\bibitem{csiszar-korner81}
I.~Csiszar and J.~K{\"o}rner, \emph{Information Theory: Coding Theorems for
  Discrete Memoryless Systems}, 2nd~ed.\hskip 1em plus 0.5em minus 0.4em\relax
  Cambridge, U.K.: Cambridge University Press, 2011.

\end{thebibliography}

\begin{IEEEbiographynophoto}{Semih Yagli}
	received his Bachelor of Science degree in Electrical and Electronics Engineering in 2013, his Bachelor of Science degree in Mathematics in 2014 both from Middle East Technical University and his Master of Arts degree in Electrical Engineering in 2016 from Princeton University.
	
	Currently, he is pursuing his Ph.D. degree in Electrical Engineering in Princeton University under the supervision of Sergio Verd\'u. His research interest include information theory, optimization, and machine learning. 
\end{IEEEbiographynophoto}

\begin{IEEEbiographynophoto}{Y\"{u}cel Altu\u{g}} 
received the B.S. and M.S. degrees in electrical and electronics engineering from Bo\u{g}azi\c{c}i University, Turkey, in 2006 and 2008, respectively and the Ph.D. degree in electrical and computer engineering from Cornell University, in 2013, where he has been awarded the ECE Director's Ph.D. Thesis Research Award. After postdoctoral appointments at Cornell University and Princeton University, he is currently a senior data scientist at Natera Inc. His research interests include Shannon theory, feedback communications, and stochastic modeling and algorithm design for next-generation DNA sequencing and genetic testing.  	
\end{IEEEbiographynophoto}

\begin{IEEEbiographynophoto}{Sergio Verd\'u}
received the Telecommunications Engineering degree from the
Universitat Polit\`{e}cnica de Barcelona in 1980, and the Ph.D. degree in Electrical Engineering from the
University of Illinois at Urbana-Champaign in 1984. Since then, he has been a member of the faculty of
Princeton University, where he is the Eugene Higgins Professor of Electrical Engineering, and is a member
of the Program in Applied and Computational Mathematics.

Sergio Verd\'{u} is  the recipient of 
the 2007 Claude E. Shannon Award, and
the 2008 IEEE Richard W. Hamming Medal. 
He is a member of both the National Academy of Engineering and the National Academy of Sciences. 
In 2016, Verd\'{u} received the National Academy of Sciences Award for Scientific Reviewing.

Verd\'{u} is a recipient of several paper awards from the IEEE: 
the 1992 Donald Fink Paper Award, 
the 1998 and 2012 Information Theory  Paper Awards, 
an Information Theory Golden Jubilee Paper Award,
the 2002 Leonard Abraham Prize Award,  
the 2006 Joint Communications/Information Theory Paper Award, 
and the 2009 Stephen O. Rice Prize from the IEEE Communications Society.  
In 1998, Cambridge University Press published his book {\em Multiuser Detection,} 
for which he received the 2000 Frederick E. Terman Award from the American Society for Engineering Education. 
He was awarded a Doctorate Honoris Causa from the Universitat  Polit\`{e}cnica de Catalunya in 2005, and was elected corresponding member of the Real Academia de Ingenier\'{i}a of Spain in 2013.

Sergio Verd\'{u} served as President of the IEEE Information Theory Society in 1997, and
on its Board of Governors (1988-1999, 2009-2014).
He has also served in various editorial capacities for the {\em IEEE Transactions on Information Theory}:
Associate Editor (Shannon Theory, 1990-1993; Book Reviews, 2002-2006),  
Guest Editor of the Special Fiftieth Anniversary Commemorative Issue
(published by IEEE Press as ``Information Theory: Fifty years of discovery"), 
and member of the Executive Editorial Board (2010-2013).
He co-chaired the  Europe-United States {\em Frontiers of Engineering} program, of
the National Academy of Engineering during 2009-2013.
He is the founding Editor-in-Chief of {\em Foundations and Trends in Communications and Information Theory}. 
Verd\'{u} served as co-chair of the 2000 and 2016 {\em IEEE International Symposia on Information Theory}.

Sergio Verd\'{u} has held visiting appointments at 
the Australian National University, 
the Technion-Israel Institute of Technology, 
the University of Tokyo, 
the University of California, Berkeley, 
the Mathematical Sciences Research Institute, Berkeley, 
Stanford University, 
and the Massachusetts Institute of Technology.
\end{IEEEbiographynophoto}

\end{document}